\documentclass[12pt,english,aps,prd]{revtex4-1}
\usepackage{CJK}
\usepackage[T1]{fontenc}
\usepackage[left=1.5cm,top=2cm,right=1.5cm,bottom=4cm]{geometry} 
\usepackage{array}
\usepackage{textcomp}
\usepackage{multirow}
\usepackage{amsmath}
\usepackage{graphicx}
\usepackage{hhline}
\usepackage{breqn}
\usepackage{comment}
\usepackage{adjustbox}

\usepackage[utf8]{inputenc} 

\usepackage{amsmath}
\usepackage{amsfonts}
\usepackage{amssymb}
\usepackage{cancel} 
\usepackage{youngtab}
\usepackage{xfrac}
\usepackage{subfigure}

\usepackage{pstricks}
\usepackage{color}

\usepackage{graphicx}
\usepackage{feynmp} 
\usepackage{subfigure} 
\usepackage{float} 

\begin{document}

\title{A non-universal $U(1)_{X}$ gauge extension to the MSSM}
\author{
J.S. Alvarado} \thanks{jsalvaradog@unal.edu.co}
\author{
Carlos E. Diaz} \thanks{cediazj@unal.edu.co}, 
\author{
R. Martinez. }
\thanks{remartinezm@unal.edu.co}
\affiliation{Departamento de F\'{i}sica$,$ Universidad Nacional de Colombia\\
 Ciudad Universitaria$,$ K. 45 No. 26-85$,$ Bogot\'a D.C.$,$ Colombia}
\date{\today}

\begin{abstract}
We propose a supersymmetric extension of the anomaly-free and three families nonuniversal $U(1)$ model, with the inclusion of four Higgs doublets and four Higgs singlets. The quark sector is extended by adding three exotic quark singlets, while the lepton sector includes two exotic charged lepton singlets, three right-handed neutrinos and three sterile Majorana neutrinos to obtain the fermionic mass spectrum. By implementing an additional $\mathbb{Z}_2$ symmetry, the Yukawa coupling terms are suited in such a way that the fermion mass hierarchy is obtained without fine-tuning. The effective mass matrix for SM neutrinos is fitted to current neutrino oscillation data to check the consistency of the model with experimental evidence, obtaining that the normal-ordering scheme is preferred over the inverse ones. The electron and up, down and strange quarks are massless at tree level, but they get masses through radiative correction at one loop level coming from the sleptons and Higgsinos contributions. We show that the model predicts a like-Higgs SM mass at electroweak scale by using the VEV according to the symmetry breaking and fermion masses. 
\end{abstract}
\maketitle 
 
\onecolumngrid
\section{Introduction}
Regardless the success of the Standard Model of electroweak interactions (SM)  \cite{SMbib} in explaining the experimental data, it is considered an incomplete model since some features remain satisfactorily unexplained. Among them, there is the fermion mass hierarchy problem as well as naturalness  problem; both have motivated many extensions of the SM, or even complete new theories. In the case of supersymmetry, it is the model which best explains the Higgs mass naturalness thanks to the exact cancellation of the quadratic divergences between contributions of particles and superpartners in the Higgs mass radiative corrections, providing  a finite mass value for the particle. 

When considering the  Minimal Supersymmetric Standard Model (MSSM) superpotential, there exists a mass-like parameter for the bilinear superfields coupling called $\mu$, which is responsible of the Higgs and Higgsino masses. This parameter is expected to be at the order of SUSY breaking scale to provide Higgsino masses. However, the lightest Higgs mass is at the electroweak scale. Thus, it can not provide the correct neutralino masses and a phenomenological Higgs mass in accordance with the data presented by ATLAS and CMS experiments \cite{Higgsobservationbib}. Additional to the above, the unexplained origin of this kind of coupling is what constitutes the $\mu$-problem \cite{muproblembib}. 

The Next to Minimal Supersymmetric Models (NMSSM) present an elegant solution to this problem \cite{NMSSMbib} by introducing new scalar singlet field. Consequently, trilinear couplings $\hat{\chi}\hat{\phi} \hat{\phi}$ can be generated in such a way that a bilinear term, $\mu\hat{\phi}\hat{\phi}$, arises when the singlet scalar field acquires a vacuum expectation value (VEV) at the SUSY breaking scale. On the other hand, when including new fields in the theory, the Higgs and Higgsinos mass matrices are changed through new coupling constants, allowing the explanation of these masses in accordance to experimental data or collider constraints. 

Looking back to the MSSM, it is known that the lightest Higgs mass can be approximated to $m_{h}^{2}\approx m_{Z}^{2}\cos^{2}2\beta + \Delta m_{h}^{2}$, where $\Delta m_{h}^{2}$ comes from the 1-loop corrections due to the top quark and stops contributions \cite{HiggsmassMSSMbib}. Then, if we consider a big value for $\tan\beta$, $\Delta m_{h}^{2}$ must be at the same order of the tree level contribution, and stop particles should have a big mass values in order to get a $125$ GeV Higgs mass. Nevertheless, in extensions of the MSSM the radiative corrections due to stop particles wouldn't be necessary for explaining the $\Delta m_{h}^{2}$ term. This may come from a seesaw mechanism that creates an explicit dependence on the scalar singlet VEV \cite{SeesawHiggsMass}. Furthermore, if the scalar singlets come as part of a $U(1)_{X}$ extended gauge symmetry (USSM) \cite{USSMbib}, the respective D-term may give a new contribution to the lightest mass at tree level, causing the new mass eigenstate to be sharing the functional form of a SM-like Higgs boson.  

The extensions of MSSM have other motivations. There are more scalar particles which can be searched in collider experiments, for instance a dark matter candidate \cite{DMSUSYbib}. Likewise, they can explain the small deviation of Higgs couplings to fermions, which turn out to be proportional to particle masses, as it has been found in experiments for tau lepton, top and bottom quarks. 

The MSSM is a two Higgs doublets anomaly free theory, where the different hypercharge values allows to each Higgs doublet to couple with different kind of quarks, forbidding flavor changing neutral currents (FCNC). In order to extend the scalar doublets content without inducing any chiral anomaly, the minimun amount of them would be four. However, while a Yukawa linear combination can be diagonalized through a rotation,  making it proportional to the particle mass, the other linear combination wouldn't be diagonal, generating then the FCNC \cite{FCNCbib}. On the other hand, from the LHC it is known the upper bound for the $tch$ vertex \cite{dataparticle} which can be explained, as new physics at tree level, from a model with multiple Higgs doublets. For this reason a SUSY theory with FNCN is still phenomenological relevant. 

In the present work, it is done a SUSY extension of the three family $U(1)_{X}$ anomaly free model \cite{nosusybib}. The non-supersymmetric model can explain the fermion mass hierarchy, as well as mixing angles for the CKM \cite{KM} and PMNS matrices \cite{CKMPMNS} just by using two Higgs doublets and a scalar singlet field which breaks the $U(1)_{X}$ symmetry giving masses to exotic particles. A singlet scalar field without VEV is required for giving masses to light fermions. Then for the corresponding SUSY extension, the scalar sector has to be doubled with different $X$-charge in such a way that the anomaly induced by higgsinos are canceled and the model would be anomaly free. After symmetry breaking, the mass matrices for the scalar, vector and fermion sector are constructed. Furthermore, from the scalar CP-even mass matrix it is found that the theory is compatible with a $125$ GeV mass for the lightest scalar particle, which we identify as the discovered Higgs boson in LHC. When considering the scalar CP-odd mass matrix, two would-be Goldstone bosons associated to the $Z$ and $Z'$ particles are found. The rest of mass eigenstates are found to be above the electroweak scale. Likewise, from the charged scalar bosons another would-be Goldstone boson is found, associated to the $W^{\pm}$ gauge bosons. In the present model, the masses of electron, quark up and quark down are zero. Then we consider the SUSY contributions to the self energies in order to generate the masses at one loop level.

\section{General Remarks of the Model}

The non-supersymmetric version of the model gives a scenario for solving the fermion mass hierarchy problem (FMH) with no need of unpleasant fine tunnings on the Yukawa coupling constants \cite{jerarquia}. The way that such problem is addressed relies in having two Higgs doublets $\Phi_{1}$ and $\Phi_{2}$; the FMH is understood partially from the VEV hierarchy among the two doublets. Also, with the help of the set configuration of the $U(1)_X$ charges for all particles, the couplings allowed by the gauge symmetry give a natural scenario for exhibiting the FMH. The inclusion of a parity symmetry $\mathbb{Z}_{2}$ helps in avoiding Yukawa terms in the Lagrangian that spoil the natural scenario wanted \cite{anzat}. The new gauge symmetry extension comes in general with  chiral anomalies, which have to be canceled in order to guarantee the renormalizability of the theory. In the model found in the literature \cite{nosusybib}, it was done by choosing the set configuration of $U(1)_X$ charges  for all fermions \cite{corrienteadicional}, such that the following anomaly equations were canceled:
\begin{widetext}
\begin{eqnarray}
\label{eq:Chiral-anomalies}
\left[\mathrm{\mathrm{SU}(3)}_{C} \right]^{2} \mathrm{\mathrm{U}(1)}_{X} \rightarrow & A_{C} &= \sum_{Q}X_{Q_{L}} - \sum_{Q}X_{Q_{R}}	\nonumber	\\
\left[\mathrm{\mathrm{SU}(2)}_{L} \right]^{2} \mathrm{\mathrm{U}(1)}_{X} \rightarrow & A_{L}  &= \sum_{\ell}X_{\ell_{L}} + 3\sum_{Q}X_{Q_{L}}	\nonumber	\\
\left[\mathrm{\mathrm{U}(1)}_{Y} \right]^{2}   \mathrm{\mathrm{U}(1)}_{X} \rightarrow & A_{Y^{2}}&=
	\sum_{\ell, Q}\left[Y_{\ell_{L}}^{2}X_{\ell_{L}}+3Y_{Q_{L}}^{2}X_{Q_{L}} \right]\nonumber	
	- \sum_{\ell,Q}\left[Y_{\ell_{R}}^{2}X_{L_{R}}+3Y_{Q_{R}}^{2}X_{Q_{R}} \right]	\nonumber	\\
\mathrm{\mathrm{U}(1)}_{Y}   \left[\mathrm{\mathrm{U}(1)}_{X} \right]^{2} \rightarrow & A_{Y}&=
	\sum_{\ell, Q}\left[Y_{\ell_{L}}X_{\ell_{L}}^{2}+3Y_{Q_{L}}X_{Q_{L}}^{2} \right]\nonumber	
	- \sum_{\ell, Q}\left[Y_{\ell_{R}}X_{\ell_{R}}^{2}+3Y_{Q_{R}}X_{Q_{R}}^{2} \right]	\nonumber	\\
\left[\mathrm{\mathrm{U}(1)}_{X} \right]^{3} \rightarrow & A_{X}&=
	\sum_{\ell, Q}\left[X_{\ell_{L}}^{3}+3X_{Q_{L}}^{3} \right]	
	- \sum_{\ell, Q}\left[X_{\ell_{R}}^{3}+3X_{Q_{R}}^{3} \right] 	\nonumber	\\	
\left[\mathrm{Grav} \right]^{2}   \mathrm{\mathrm{U}(1)}_{X} \rightarrow & A_{\mathrm{G}}&=
	\sum_{\ell, Q}\left[X_{\ell_{L}}+3X_{Q_{L}} \right]
	- \sum_{\ell, Q}\left[X_{\ell_{R}}+3X_{Q_{R}} \right],
\end{eqnarray}
\end{widetext}
\noindent
\onecolumngrid
where subscripts $Q$ and $l$ account for quarks and leptons, respectively. Moreover, subscripts $L$ and $R$ correspond to left-handed and right-handed chiralities, respectively. Exotic fermions were also included in the model for accomplishing a free anomaly model, in the way that new degrees of freedom enter into the equation (\ref{eq:Chiral-anomalies}). Thus, there is a bigger set of $U(1)_X$ charges than the SM particles for fulfilling both anomaly cancellation and FMH. In the quark sector, an up-like quark $\mathcal{T}$ and two down-like quarks, $\mathcal{J}^1$ and $\mathcal{J}^2$, come into the bargain. The additional particles in the lepton sector are two charged leptons, $E$ and $\mathcal{E}$; three Dirac right handed neutrinos, $\nu^e_{R}$, $\nu^\mu_{R}$ and $\nu^\tau_{R}$; and three Majorana neutrinos $\mathcal{N}_{R}^{1,2,3}$. The majorana particles do not contribute to the anomaly equations, but they were included for giving masses to neutrinos through inverse seesaw mechanism (ISS) \cite{ISSbib}, according to neutrino oscillation experiments which give information about squared mass differences and mixing angles.
For breaking the new symmetry into the SM gauge symmetry, an scalar singlet $\chi$ was added with a VEV around the TeV scale. Therefore, the model contains the following spontaneous symmetry breaking chain:
\begin{eqnarray}
    \mathrm{SU(3)}_{C}\otimes
    \mathrm{SU(2)}_{L}\otimes 
    \mathrm{U(1)}_{Y} \otimes 
    \mathrm{U(1)}_{X} &\overset{\chi}{\longrightarrow}&\\
    \mathrm{SU(3)}_{C}\otimes
    \mathrm{SU(2)}_{L}\otimes 
    \mathrm{U(1)}_{Y} &\overset{\Phi}{\longrightarrow} &
    \mathrm{SU(3)}_{C}\otimes
    \mathrm{U(1)}_{Q}. \nonumber
\end{eqnarray}

Because the lightest fermions, electron, down quark and up quark, did not acquire masses at tree level, another scalar singlet $\sigma$ had to be included for giving masses to such particles through radiative corrections. However, in the SUSY version of the model this scalar field is still present but now the corresponding superpartners of the particles inside the loop represent an additional SUSY contribution to their masses.

For the minimal supersymmetric extension, all fields are upgraded to superfields; we denote a superfield with a hat symbol, as usual.  The number of scalar particles has to be doubled in comparison to the non-SUSY version, otherwise the model would be anomalous due to Higgsinos. Therefore, new added fields are $\hat{\Phi}_{1}'$, $\hat{\Phi}_{2}'$, $\hat{\chi}'$ and $\sigma^{\prime}$. The introduced scalar fields have the same hypercharge and X charges as the non-primed partners, but with opposite sign to secure anomaly cancellation. For getting the right masses of the gauge bosons in the SM, the following condition must be imposed on the VEVs of the Higgs doublets,
\begin{equation}
\sqrt{v_{1}^{2}+v_{2}^{2}+v_{1}^{\prime2}+v_{2}^{\prime 2}}=v=246 GeV.    \label{VEVcondition}
\end{equation}

The bosonic and fermonic content of the model explored in this paper is shown in the tables \ref{tab:Bosonic-content-A-B} and \ref{tab:Particle-content-A-B}, respectively.

\section{Scalar and gauge boson Sector}

\begin{table}
\caption{Scalar content of the model, non-universal $X$ quantum number, $\mathbb{Z}_{2}$ parity and hypercharge}
\label{tab:Bosonic-content-A-B}
\centering
\begin{tabular}{lll cll}\hline\hline 
\multirow[l]{3}{*}{
\begin{tabular}{l}
    Higgs    \\
    Scalar  \\
    Doublets
\end{tabular}
}
&\multicolumn{2}{l}{}&
\multirow[l]{3}{*}{
\begin{tabular}{l}
    Higgs    \\
    Scalar  \\
    Singlets
\end{tabular}
} 
&\multicolumn{2}{l}{}\\ 
 &&&
 && \\ 
 &$X^{\pm}$&$Y$&
 &$X^{\pm}$&$Y$
\\ \hline\hline 
$\small{\hat{\Phi}_{1}=\begin{pmatrix}\hat{\phi}_{1}^{+}\\\frac{\hat{h}_{1}+v_{1}+i\hat{\eta}_{1}}{\sqrt{2}}\end{pmatrix}}$&$\sfrac{+2}{3}^{+}$&$+1$&
$\hat{\chi}=\frac{\hat{\xi}_{\chi}+v_{\chi}+i\hat{\zeta}_{\chi}}{\sqrt{2}}$	&	$\sfrac{-1}{3}^{+}$	&	$0$	\\
$\small{\hat{\Phi}_{2}=\begin{pmatrix}\hat{\phi}_{2}^{+}\\\frac{\hat{h}_{2}+v_{2}+i\hat{\eta}_{2}}{\sqrt{2}}\end{pmatrix}}$&$\sfrac{+1}{3}^{-}$&$+1$& $\sigma=\frac{\hat{\sigma}_{\chi}+i\hat{\zeta}_{\sigma}}{\sqrt{2}} $ &$ \sfrac{-1}{3}^{-} $ & $ 0 $		\\
$\small{\hat{\Phi}^\prime_{1}=\begin{pmatrix}\frac{\hat{h}_{1}'+v_{1}'+i\hat{\eta}_{1}'}{\sqrt{2}}\\\hat{\phi}_{1}^{-\prime}\end{pmatrix}}$&$\sfrac{-2}{3}^{+}$&$-1$&
$\hat{\chi}'=\frac{\hat{\xi}'_{\chi}+v_{\chi}'+i\hat{\zeta}'_{\chi}}{\sqrt{2}}$	&	$\sfrac{+1}{3}^{+}$ &	0\\
$\small{\hat{\Phi}^\prime_{2}=\begin{pmatrix}\frac{\hat{h}_{2}'+v_{2}'+i\hat{\eta}_{2}'}{\sqrt{2}}\\\hat{\phi}_{2}^{-\prime}\end{pmatrix}}$&$\sfrac{-1}{3}^{-}$&$-1$& $\sigma^{\prime} = \frac{\hat{\xi}_{\sigma}^{\prime}+i\hat{\zeta}_{\sigma}^{\prime}}{\sqrt{2}}$ & $\sfrac{+1}{3}^{-}$ &	0	\\\hline\hline
\end{tabular}
\end{table}
The Lagrangian for the scalar sector  that describes the minimal supersymetric extension to the $U(1)_{X}$ model in the literature is given by the addition of a F-terms potential, a D-terms potential and a soft-supersymetry breaking potential. The F-terms were obtained from the following superpotential:
\begin{align}
    W_{\phi}&=-\mu_{1}\hat{\Phi}'_{1}\hat{\Phi}_{1}-\mu_{2}\hat{\Phi}'_{2}\hat{\Phi}_{2} - \mu_{\chi}\hat{\chi} '\hat{\chi} - \mu_{\sigma}\hat{\sigma} '\hat{\sigma} + \lambda_{1}\hat{\Phi}_{1}^{\prime}\hat{\Phi}_{2}\hat{\sigma}^{\prime} + \lambda_{2}\hat{\Phi}_{2}^{\prime}\hat{\Phi}_{1}\sigma. 
\end{align}
which is obtained according to the symmetry properties of the scalar superfields given in Table I. Thus, the F-terms potential for scalar fields reads

\begin{small}
\begin{align}
V_{F}&=\mu_{1}^2(\Phi_{1}^\dagger\Phi_{1}+\Phi_{1}^{\prime\dagger}\Phi_{1}^{\prime})+\mu_{2}^2(\Phi_{2}^\dagger\Phi_{2}+\Phi_{2}^{\prime\dagger}\Phi_{2}^{\prime})
+\mu_{\chi}^2(\chi^*\chi+\chi^{\prime*}\chi') + +\mu_{\sigma}^2(\sigma^*\sigma+\sigma^{\prime*}\sigma^{\prime})\nonumber\\
&+ \Big(\lambda_{1}^{2}|\epsilon_{ij}\Phi_{1}^{\prime i}\Phi_{2}^{j}|^{2} + \lambda_{2}^{2}|\epsilon_{ij}\Phi_{2}^{\prime i}\Phi_{1}^{j}|^{2}  + \lambda_{1}^{2}( \Phi_{2}^{\dagger} \Phi_{2} + \Phi_{1}^{\prime \dagger}\Phi_{1}^{\prime}\sigma^{\prime *}\sigma^{\prime} + \lambda_{2}^{2}( \Phi_{1}^{\dagger}\Phi_{1} + \Phi_{2}^{\prime \dagger}\Phi_{2}^{\prime})\sigma^{*}\sigma  -\lambda_{1}\mu_{1}\Phi_{1}^{\dagger}\Phi_{2}\sigma^{\prime} \nonumber\\
&- \lambda_{1}\mu_{2}\Phi_{2}^{\prime \dagger }\Phi_{1}^{\prime}\sigma^{\prime}  -\lambda_{2}\mu_{1}\Phi_{1}^{\prime \dagger}\Phi_{2}^{\prime}\sigma -\lambda_{2}\mu_{2}\Phi_{2}^{\dagger}\Phi_{1}\sigma - \lambda_{1}\mu_{\sigma}\epsilon_{ij}\Phi_{1}^{\prime i}\Phi_{2}^{j} -\lambda_{2}\mu_{\sigma}\epsilon_{ij}\Phi_{2}^{\prime i }\Phi_{1}^{j} + h.c. \Big )
\end{align}
\end{small}

On the other hand the D-terms potential, consequence of gauge symmetry, turns out to be
\begin{widetext}
\begin{align}
V_{D}&=\frac{g^{2}}{2}\Big[ |\Phi_{1}^{\dagger}\Phi_{2}|^{2}+|\Phi_{1}^{\prime\dagger}\Phi_{2}'|^2+|\Phi_{1}^{\prime\dagger}\Phi_{1}|^2+|\Phi_{1}^{\prime\dagger}\Phi_{2}|^2+|\Phi_{2}^{\prime\dagger}\Phi_{1}|^2+|\Phi_{2}^{\prime\dagger}\Phi_{2}|^2-|\Phi_{1}|^{2}|\Phi_{2}|^{2}-|\Phi_{1}^{\prime}|^{2}|\Phi_{2}^{\prime}|^{2} \Big]\nonumber\\
 &+\frac{g^{2} + g^{\prime 2}}{8}(\Phi_{1}^{\dagger}\Phi_{1}+\Phi_{2}^{\dagger}\Phi_{2}-\Phi_{1}^{\prime\dagger}\Phi_{1}^{\prime}-\Phi_{2}^{\prime\dagger}\Phi_{2}^{\prime})^{2} \nonumber\\
 &+\frac{g_{X}^{2}}{2}\left[\frac{2}{3}(\Phi_{1}^{\dagger}\Phi_{1}-\Phi_{1}^{\prime\dagger}\Phi_{1}^{\prime})+\frac{1}{3}(\Phi_{2}^{\dagger}\Phi_{2}-\Phi_{2}^{\prime\dagger}\Phi_{2}^{\prime})-\frac{1}{3}(\chi^{*}\chi-\chi^{\prime*}\chi^{\prime}) -\frac{1}{3}(\sigma^{*}\sigma-\sigma^{\prime*}\sigma^{\prime})\right]^{2}
\end{align}
\noindent
where the last term corresponds to the D-term associated to the $U(1)_{X}$ gauge symmetry and has a very important role for giving the lightest Higgs boson a mass around $125$GeV. This will be treated later. Finally, the soft supersymmetry breaking potential turns out to be:
\begin{small}\begin{align}\label{soft}
    V_{soft}&=-m_{1}^{2}\Phi_{1}^{\dagger}\Phi_{1} - {m}_{1}^{\prime 2}{\Phi}_{ 1}^{\prime \dagger}\Phi'_{1} - m_{2}^{2}\Phi_{2}^{\dagger}\Phi_{2} - {m}_{2}^{\prime 2}\Phi _{2}^{\prime\dagger}\Phi'_{2}-m_{\chi}^{2}\chi^{\dagger}\chi - {m}_{\chi}^{\prime 2}{\chi}^{\prime\dagger}\chi'-m_{\sigma}^{2}\sigma^{\dagger}\sigma - {m}_{\sigma}^{\prime 2}{\sigma}^{\prime\dagger}\sigma'  \nonumber\\
    &+\bigg[\mu_{11}^{2}\epsilon_{ij}({\Phi}_{1}^{\prime i}\Phi_{1}^{j}) +\mu_{22}^{2}\epsilon_{ij}({\Phi}_{2}^{\prime i}\Phi_{2}^{j}) +\mu_{\chi\chi}^{2}(\chi\chi') +\mu_{\sigma\sigma}^{2}(\sigma\sigma') - \tilde{\lambda}_{1}\Phi_{1}^{\prime \dagger}\Phi_{2}\sigma^{\prime} - \tilde{\lambda}_{2}\Phi_{2}^{\prime \dagger}\Phi_{1}\sigma \nonumber\\
    &+ \frac{2\sqrt{2}}{9}(k_{1}\Phi_{1}^{\dagger}\Phi_{2}\chi' -k_{2}\Phi_{1}^{\dagger}\Phi_{2}\chi^*+k_{3}\Phi_{1}'{}^{\dagger}\Phi_{2}'\chi -k_{4}\Phi_{1}'{}^{\dagger}\Phi_{2}'\chi'{}^*)+h.c.\bigg]
\end{align}
\end{small}
\end{widetext}
where the last terms, proportional to the coupling constants named $k_{1},k_{2},k_{3}$ and $k_{4}$, break also softly the parity symmetry. These trilinear terms avoid the massless feature of some scalar particles, as we will show later. It is important to mention that the soft supersymmetry breaking potential also includes  bilinear terms for sfermions and gauginos. Nonetheless, since those terms are not required for our calculations we have decided that it is not necessary to present them in the current work. Then, by adding all contributions, the scalar potential for the Higgs bosons reads
\begin{equation}\label{V}
V_{H}=V_{F}+V_{D}+V_{soft} .   
\end{equation}
When considering the potential $V_{F}$ it can be seen that, before including soft SUSY breaking, particles within the fields $\Phi_{i}$ and $\Phi_{i}^{\prime}$, i=1,2, are expected to have the same mass $\mu_{i}$ due to the absence of mixing terms. Then, with the inclusion of a soft breaking potential, $m_i^2$, $m_i^{\prime2}$ terms arise and the diagonal entries change according to the effective parameters  $m_{Hk}^{2}=m_{k}^{2}+\mu_{k}^{2}$ and $m_{Hk}^{\prime2}=m_{k}^{\prime2}+\mu_{k}^{2}$, ensuring that different Higgs doublets have now different mass eigenvalues. This is also exhibited by the scalar singlets, where diagonal entries are written in terms of the effective parameters $M_{\chi}^{2}=m_{\chi}^{2}+\mu_{\chi}^{2}$ and $M_{\chi}^{\prime2}=m_{\chi}^{\prime2}+\mu_{\chi}^{2}$. The following minima conditions for the Higgs potential have to be fulfilled:
\begin{align}
m_{H1}^{2}+\frac{1}{8}(g^{2}+g'{}^{2})C_{EW}+\frac{g_{X}^{2}}{9}c_{X}-\mu_{11}\frac{v_{1}^{\prime}}{v_{1}} + \frac{\lambda_{2}^{2}}{2}v_{2}^{\prime 2}&=0\nonumber \\
m_{H1}^{\prime 2}-\frac{1}{8}(g^{2}+g'{}^{2})C_{EW}-\frac{g_{X}^{2}}{9}c_{X}-\mu_{11}\frac{v_{1}}{v_{1}^{\prime}}+ \frac{\lambda_{1}^{2}}{2}v_{1}^{\prime 2}&=0\nonumber \\
m_{H2}^{2}+\frac{1}{8}(g^{2}+g'{}^{2})C_{EW}+\frac{g_{X}^{2}}{18}c_{X}-\mu_{22}\frac{v_{2}^{\prime}}{v_{2}}+ \frac{\lambda_{1}^{2}}{2}v_{2}^{ 2}&=0\nonumber \\
m_{H2}^{\prime 2}-\frac{1}{8}(g^{2}+g'{}^{2})C_{EW}-\frac{g_{X}^{2}}{18}c_{X}-\mu_{22}\frac{v_{2}}{v_{2}^{\prime}}+ \frac{\lambda_{2}^{2}}{2}v_{1}^{ 2}&=0\nonumber \\
M_{\chi}^{2}-\frac{g_{X}^{2}}{18}c_{X}-\mu_{\chi\chi}\frac{v_{\chi}^{\prime}}{v_{\chi}}&=0\nonumber \\
M_{\chi}^{\prime 2}+\frac{g_{X}^{2}}{18}c_{X}-\mu_{\chi\chi}\frac{v_{\chi}}{v_{\chi}^{\prime}}&=0\label{minimaconditions}, 
\end{align}
where we also defined $C_{EW}=v_{1}^{2}+v_{2}^{2}-v_{1}^{\prime 2}-v_{2}^{\prime 2}$ and $C_{X}=2v_{1}^{2}+v_{2}^{2}-2v_{1}^{\prime 2}-v_{2}^{\prime 2}+v_{\chi}^{\prime 2}-v_{\chi}^{2}$.

\subsection{CP-even masses}

Taking the VEV for all scalar fields, we get the mass matrices for the different Higgs boson particles, that also respect the minima conditions, eq. (\ref{minimaconditions}). The 125 GeV Higgs boson is a CP-even scalar, and it must be obtained from the diagonalization of the following $6\times 6$ mass matrix in the $(h_{1},h_{1}',h_{2},h_{2}',\xi_{\chi},\xi'_{\chi},\xi_{\sigma},\xi'_{\sigma})$ basis:
\begin{align}\label{h}
    \frac{1}{2}M_{h}^{2}&=\begin{pmatrix}
    M_{hh} & M_{h\xi}  \\
    M_{h\xi}^{T} & M_{\xi\xi} 
    \end{pmatrix}.
\end{align}
$M_{hh}$ is a $4\times 4$ matrix accounting for the mixings among the CP-even fields of the doublets in the model, and it reads
\begin{widetext}
\begin{align}\label{mphi}
M_{hh}&=\begin{pmatrix}
    f_{4g}v_{1}^{2}-\frac{v_{2}f_{1k}}{9v_{1}}+\frac{v_{1}'\mu_{11}^{2}}{2v_{1}} & -f_{4g}v_{1}v_{1}'-\frac{\mu_{11}^{2}}{2} &  f_{2g}v_{1}v_{2}+\frac{f_{1k}}{9}&-f_{2g}v_{1}v_{2}'+ \frac{1}{2}\lambda_{2}^{2}v_{1}v_{2}^{\prime}\\
    * & f_{4g}v_{1}'{}^{2}-\frac{v_{2}'f_{2k}}{9v_{1}'}+\frac{v_{1}\mu_{11}^{2}}{2v_{1}'}& -f_{2g}v_{1}'v_{2} + \frac{1}{2}\lambda_{1}^{2}v_{2}v_{1}^{\prime}&f_{2g}v_{1}'v_{2}'+\frac{f_{2k}}{9}\\
    *&*& f_{1g}v_{2}^{2}-\frac{v_{1}f_{1k}}{9v_{2}}+\frac{v_{2}'\mu_{22}^{2}}{2v_{2}}& -f_{1g}v_{2}v_{2}'-\frac{\mu_{22}^{2}}{2}\\
    *&*&*&f_{1g}v_{2}'{}^{2}-\frac{v_{1}'f_{2k}}{9v_{2}'}+\frac{v_{2}\mu_{22}^{2}}{2v_{2}'}
    \end{pmatrix}.
\end{align}
\end{widetext}
\onecolumngrid
\noindent
The mixings between scalar doublets and singlets are written in the $4\times 4$ $M_{h\xi}$  matrix and it is given by:
\small
\begin{align}
     M_{h\xi}&=\begin{pmatrix}
    \frac{1}{9}(k_{2}v_{2}-g_{X}^2v_{1}v_{\chi}) & \frac{1}{9}(-k_{1}v_{2}+g_{X}^2v_{1}v_{\chi}') & \frac{1}{2\sqrt{2}}(\tilde{\lambda_{2}}v_{2}^{\prime} - \lambda_{2}\mu_{2}v_{2}) & -\frac{1}{2\sqrt{2}}(\lambda_{1}\mu_{1}v_{2} + \lambda_{2}\mu_{\sigma}v_{2}^{\prime} ) \\
    \frac{1}{9}(-k_{3}v_{2}'+g_{X}^2v_{1}'v_{\chi}) & \frac{1}{9}(k_{4}v_{2}'-g_{X}^2v_{1}'v_{\chi}') -&  -\frac{1}{2\sqrt{2}}(\lambda_{2}\mu_{1}v_{2}^{\prime} + \lambda_{1}\mu_{\sigma}v_{2}) & \frac{1}{2\sqrt{2}}(\tilde{\lambda_{1}}v_{2} - \lambda_{1}\mu_{2}v_{2}^{\prime})\\
    \frac{1}{9}(k_{2}v_{1}-\frac{1}{2}g_{X}^2v_{2}v_{\chi})& \frac{1}{9}(-k_{1}v_{1}+\frac{1}{2}g_{X}^2v_{2}v_{\chi}') &  -\frac{1}{2\sqrt{2}}(\lambda_{2}\mu_{2}v_{1} + \lambda_{1}\mu_{\sigma}v_{1}^{\prime}) & \frac{1}{2\sqrt{2}}(\tilde{\lambda_{1}}v_{1}^{\prime} - \lambda_{1}\mu_{1}v_{1})\\
    \frac{1}{9}(-k_{3}v_{1}'+\frac{1}{2}g_{X}^2v_{2}'v_{\chi})& \frac{1}{9}(k_{4}v_{1}'-\frac{1}{2}g_{X}^2v_{2}'v_{\chi}') & \frac{1}{2\sqrt{2}}(\tilde{\lambda_{2}}v_{1} - \lambda_{2}\mu_{1}v_{1}^{\prime}) & -\frac{1}{2\sqrt{2}}(\lambda_{1}\mu_{2}v_{1}^{\prime} + \lambda_{2}\mu_{\sigma}v_{1})
    \end{pmatrix}
\end{align}
\normalsize
\noindent
Finally, the mixing matrix between Higgs singlets, $M_{\xi \xi}$, is written in the following equation
\footnotesize
\begin{align}\label{11}
    M_{\xi\xi}&=
    \begin{pmatrix}
   \frac{g_{X}^2}{18}v_{\chi}^2 +\frac{v_{\chi}'\mu_{\chi\chi}^2}{2v_{\chi}}-\frac{k_{23}}{9v_{\chi}} & -\frac{g_{X}^2}{18}v_{\chi}v_{\chi}'-\frac{\mu_{\chi\chi}^{2}}{2} & 0 & 0\\
    * &  \frac{g_{X}^2}{18}v_{\chi}'{}^2 +\frac{v_{\chi}\mu_{\chi\chi}^2}{2v_{\chi}'}-\frac{k_{14}}{9v_{\chi}'} & 0 & 0 \\
    * & * &M_{\sigma}^{2} +\frac{\lambda_{2}^{2}}{4}(v_{1}^{2}+v_{2}^{\prime 2})& -\frac{\mu_{\sigma\sigma}}{2} \\
    * & * & * &M_{\sigma}^{\prime 2} +\frac{\lambda_{1}^{2}}{4}(v_{2}^{2}+v_{1}^{\prime 2}) 
    \end{pmatrix} 
\end{align} 
\normalsize
\noindent

Aimed in giving shorter expressions we have defined the coefficients  $f_{ng}=\frac{g^{2}+g'{}^{2}}{8}+\frac{n}{18}g_{X}^{2}$ with n an integer, $f_{1k}=k_{2}v_{\chi}-k_{1}v_{\chi}'$, $f_{2k}=-k_{3}v_{\chi}+k_{4}v_{\chi}'$, $k_{23}=k_{2}v_{1}v_{2}-k_{3}v_{1}'v_{2}'$ and $k_{14}=-k_{1}v_{1}v_{2}+k_{4}v_{1}'v_{2}'$ and the sigma masses $M_{\sigma}=\frac{1}{2}(\mu_{\sigma}^{2} + m_{\sigma}^{2}) - \frac{g_{X}^{2}}{36}(2v_{1}^{2}+v_{2}^{2}-2v_{1}^{\prime 2}-v_{2}^{\prime 2} - v_{\chi}^{2} +v_{\chi}^{\prime 2})$ and  $M_{\sigma}^{\prime}=\frac{1}{2}(\mu_{\sigma}^{2} + m_{\sigma}^{\prime 2}) + \frac{g_{X}^{2}}{36}(2v_{1}^{2}+v_{2}^{2}-2v_{1}^{\prime 2}-v_{2}^{\prime 2} - v_{\chi}^{2} +v_{\chi}^{\prime 2})$  \newline

In order to diagonalize the CP-even Higgs mass matrix given in equation (\ref{h}) we make use of perturbation theory by implementing a seesaw mechanism which requires then to specify a hierarchy among the parameters present in the different blocks. Firstly, we suppose their order of magnitude such that they obey $\mathcal{O}( M_{\xi\xi})\gg \mathcal{O}( M_{h\xi})\gg\mathcal{O}( M_{hh})$ which implies a hierarchy among the different parameters in the potentials (eq. (\ref{V})). To have a correct phenomenological theory we assume $\mu_{\chi\chi}, \mu_{\sigma \sigma}, M_{\sigma}, M_{\sigma}^{\prime}\gg\mu_{11}\text{, }\mu_{22} \gg k_{i}v_{j}\gg g_{X}^{2}v_{\chi}v_{j}\text{, }g_{X}^{2}v'_{\chi}v_{j}\text{, }g_{X}^{2}v_{\chi}v'_{j}\text{, }g_{X}^{2}v'_{\chi}v'_{j}$, where $i=1,2,3,4$ and $j=1,2$.  Therefore, the conditions for implementing a seesaw mechanism are fulfilled which leads to a block diagonal mass matrix made of two independent $4\times 4$  matrices. Additionally, as the rank of $M_{h}$ is 8, all mass eigenstates are massive. Considering that a Higgs singlet has not been observed in the experiments, it would be expected for them to acquire mass at high energy scale, supporting our hierarchy choice and implying that  $v_{\chi}$ and $v'_{\chi}$ should be at least at the TeV scale, thus they satisfy $v_{\chi}, v'_{\chi} \gg v_{j}, v'_{j}$, where $j=1,2$. Additional to our assumptions, a small program in C++ was written to find the parameter region in which a $125$(GeV) value match the lightest eigenvalue using a simple Monte Carlo exploration. It was found that $k_{i} \sim 10^3$, $0 < \lambda_{i}, \tilde{\lambda}_{i} < 10^{3} $, $\mu_{11},\mu_{22} > 10^{4}$, $\mu_{\chi\chi}, \mu_{\sigma \sigma}, M_{\sigma}, M_{\sigma}^{\prime} > 10^{8}$ and $ v_{\chi},v'_{\chi} > 10^{3}$, which in fact satisfy the considered parameter hierarchy. The singlet VEVs lower bound are fixed in such a way that guarantees a lightest eigenvalue coming from a scalar doublets mixture. 
\begin{align}\label{hSS}
    \Tilde{M}_{h}&\approx\begin{pmatrix}
    \Tilde{M}_{hh} & 0 \\
    0 & M_{\xi\xi}
    \end{pmatrix}, \nonumber\\  
    \Tilde{M}_{hh}&=M_{hh}-M_{h\xi}M_{\xi\xi}^{-1}M_{h\xi}^{T} \nonumber\\
    &\approx M_{hh}.
\end{align}

 Since the matrix rank for $M_{hh}$ is 4, all Higgs particles are massive and acquire their tree level masses through $M_{hh}$, then they receive small contributions coming from the seesaw rotation ($M_{h\xi}M_{\xi\xi}^{-1}M_{h\xi}^{T}$), which can be neglected due to the parameters' order of magnitude. As a result we can assume that $\Tilde{M}_{hh} \approx M_{hh}$. In this way it is possible to get an expression for the heaviest singlet Higgs particles by the diagonalization of the $4\times 4$ decoupled matrix which in fact is a $2\times 2$ block diagonal one since the singlets does not mix among them. We are considering that the heaviest particles must depend on $\mu_{\chi\chi}$, $\mu_{\sigma\sigma}$, $M_{\sigma}$ and $M_{\sigma}^{\prime}$, which are the biggest parameters in the theory; while for the two lightest of these singlets depend then on $v_{\chi}$, $v'_{\chi}$, $k_{i}$ and $\lambda_{1}$ with the condition that the sum must reconstruct the original trace. In this way we can get the tree level approximation of these eigenvalues which can be written as:

\begin{align}
    m_{h5}^{2}&\approx\frac{g_{X}^{2}}{9}(v_{\chi}^{2}+v_{\chi}'{}^{2})-\frac{2}{9}\frac{v_{1}v_{2}(k_{2}v_{\chi}'-k_{1}v_{\chi})+v_{1}'v_{2}'(k_{4}v_{\chi}-k_{3}v_{\chi}')}{v_{\chi}v_{\chi}'},\\
     m_{h6}^{2}&\approx\mu_{\chi\chi}^{2}\frac{v_{\chi}^{2}+v_{\chi}'{}^{2}}{v_{\chi}v_{\chi}'}, \nonumber
     \end{align}
 \begin{small}
     \begin{align}
     m_{h7}^{2}=(M_{\sigma}^{2}+M_{\sigma}^{\prime 2}) +  \frac{1}{4}[\lambda_{2}^{2}(v_{1}^{2}+v_{2}^{\prime 2})+ & \lambda_{1}^{2}(v_{2}^{2}+v_{1}^{\prime 2})] \nonumber \\
     &- \sqrt{\mu_{\sigma \sigma}^{4} - \left((M_{\sigma}^{2}+M_{\sigma}^{\prime 2})+\frac{1}{4}[\lambda_{2}^{2}(v_{1}^{2}+v_{2}^{\prime 2}) - \lambda_{1}^{2}(v_{2}^{2}+v_{1}^{\prime 2})]\right)^{2}}, \\
     m_{h8}^{2}=(M_{\sigma}^{2}+M_{\sigma}^{\prime 2}) +  \frac{1}{4}[\lambda_{2}^{2}(v_{1}^{2}+v_{2}^{\prime 2})+ & \lambda_{1}^{2}(v_{2}^{2}+v_{1}^{\prime 2})]  \nonumber\\
     &+ \sqrt{\mu_{\sigma \sigma}^{4} - \left((M_{\sigma}^{2}+M_{\sigma}^{\prime 2})+\frac{1}{4}[\lambda_{2}^{2}(v_{1}^{2}+v_{2}^{\prime 2}) - \lambda_{1}^{2}(v_{2}^{2}+v_{1}^{\prime 2})]\right)^{2}} \nonumber
\end{align}
\end{small}

Since the singlet $\sigma$ acquire a null vacuum expectation value there is not a minimum condition which relates $M_{\sigma}$, $M_{\sigma}^{\prime}$ and $\mu_{\sigma\sigma}$ as it happened with the other scalar particles. In fact, there is more freedom for making a hierarchy choice among them. However, this particles are expected to a reside at an unreachable energy scale for the current experiments and does not represent the main focus of this work; even though they have an important role in fermion mass generation as it will be shown later. For finding corresponding $M_{\phi}$ eigenvalues we took the following approach. We consider that at least one of the Higgs particles must be proportional to the electroweak VEVs such that it can be identified as the SM Higgs particle. Therefore, the other three must be heavy and functions of the SSB parameters and $\mu_{11}$, $\mu_{22}$ and $k_{i}$; reason for which we took a small doublet's VEV limit ($v_{1},v_{2},v_{1}^{\prime},v_{2}^{\prime}\rightarrow 0$), and it is found that the matrix rank decreases to 3. In fact, it means that there is an eigenvalue which depends entirely on the EW VEVs and it can be identified as a SM Higgs particle. Furthermore, when considering this limit we are able to get the next two heavy states by discarding the additive terms proportional to those VEV. As a result, the matrix $M_{\phi}$ reduces to the following form:
\begin{align}\label{v=0}
M_{\phi}(v_{i},v_{i}'\rightarrow 0)&=\begin{pmatrix}
\frac{\mu_{11}^2}{2}\frac{v_{1}^{\prime}}{v_{1}}    & -\frac{\mu_{11}^2}{2} &0&0\\
*                           &\frac{\mu_{11}^2}{2}\frac{v_{1}}{v_{1}^{\prime}}   &0&0\\
*                           &*          & \frac{\mu_{22}^2}{2}\frac{v_{2}^{\prime}}{v_{2}}&-\frac{\mu_{22}^2}{2}\\
*                           &*                      &*&\frac{\mu_{22}^2}{2}\frac{v_{2}}{v_{2}^{\prime}}
\end{pmatrix} 
\end{align}
\noindent
Therefore, we find two decoupled $2\times 2$ matrices with determinant equal to zero, arising the two heaviest states. Then, in the tree level the eigenvalues are given by:
\begin{align}\label{h34}
    m_{h3}^{2}&\approx \mu_{11}^{2}\frac{v_{1}^{2}+v_{1}^{\prime 2}}{v_{1}v_{1}^{\prime}}, & m_{h4}^{2}&\approx \mu_{22}^{2}\frac{v_{2}^{2}+v_{2}^{\prime 2}}{v_{2}v_{2}^{\prime}}.
\end{align}

At this point, there are only two CP-even particles for which one of them must be the like-Higgs SM. Returning to the original mass matrix $M_{hh}$ (\ref{mphi}), it is known that eigenvalues are roots of the characteristic polynomial. In our case, the polynomial is a fourth degree  one. It can be solved analytically by using a general solution given by Ferrari's method \cite{Ferraribib}. 
It is worth to notice that by taking the small VEV limit in the corresponding two general solutions, accounting for the two heavy eigenstates, the eigenvalues in eq. (\ref{h34}) are reproduced. On the other hand, the smaller value acquired by the general solution would correspond to the SM-like Higgs particle but the resulting expression for the latter is too complicated by using this method, requiring then a different approach for obtain it. However, by taking again the small VEV limit mentioned before and taking into account the chosen hierarchy, now the dominant terms are proportional to $k_{i}$ and it results in the following expression. 
\begin{align}\label{h2}
m_{h2}^{2}&\approx \frac{2v^{2}(v_{1}v_{2}(k_{1}v_{\chi}^{\prime}-k_{2}v_{\chi})+v_{1}^{\prime}v_{2}^{\prime}(k_{3}v_{\chi}-k_{4}v_{\chi}^{\prime}))}{9(v_{1}^{2}+v_{1}^{\prime 2})(v_{2}^{2}+v_{2}^{\prime 2})}.
\end{align}

Finally we focus on the lightest CP-even Higgs particle, which so far in our approximation became massless, but must match the $125$ GeV observed one. For starting, we considered the determinant of the $4\times 4$ matrix given by eq. (\ref{mphi}) only taking into account the dominant terms which are proportional to $\mu_{11}^{2}\mu_{22}^{2}$. It reads:

\begin{align}
Det(\Tilde{M}_{hh}) &\approx \frac{\mu_{11}^{2}\mu_{22}^{2}}{2592}\Bigg[(k_{3}v_{\chi}-k_{4}v_{\chi}^{\prime})\bigg((9(g^{2}+g^{\prime 2})+16g_{X}^{2})\frac{(v_{1}^{2}-v_{1}^{\prime 2})^{2}}{v_{1}v_{2}}+(9(g^{2}+g^{\prime 2})+4g_{X}^{2})\frac{(v_{2}^{2}-v_{2}^{\prime 2})^{2}}{v_{1}v_{2}}\nonumber \\
&+2(9(g^{2}+g^{\prime 2})+8g_{X}^{2})\frac{(v_{1}^{2}-v_{1}^{\prime 2})(v_{2}^{2}-v_{2}^{\prime 2})}{v_{1}v_{2}}\bigg) +(k_{1}v_{\chi}^{\prime}-k_{2}v_{\chi})\bigg((9(g^{2}+g^{\prime 2})+16g_{X}^{2})\frac{(v_{1}^{2}-v_{1}^{\prime 2})^{2}}{v_{1}^{\prime}v_{2}^{\prime}}\nonumber\\
&+(9(g^{2}+g^{\prime 2})+4g_{X}^{2})\frac{(v_{2}^{2}-v_{2}^{\prime 2})^{2}}{v_{1}^{\prime}v_{2}^{\prime}}+2(9(g^{2}+g^{\prime 2})+8g_{X}^{2})\frac{(v_{1}^{2}-v_{1}^{\prime 2})(v_{2}^{2}-v_{2}^{\prime 2})}{v_{1}^{\prime}v_{2}^{\prime}}\bigg)\Bigg]
\end{align}
\onecolumngrid
The lightest Higgs mass eigenvalue is found by dividing this determinant by the other three found mass eigenvalues given by equations (\ref{h34})-(\ref{h2}), which can be written as
\footnotesize
\begin{align}\label{Higgsobservedmass}
m_{h1}^{2}&\approx \frac{g_{X}^{2}(2v_{1}^{2}+v_{2}^{2}-2v_{1}^{\prime 2}-v_{2}^{\prime 2})^{2} }{9(v_{1}^{2}+v_{2}^{2}+v_{1}^{\prime 2}+v_{2}^{\prime 2})}+ \frac{ (g^{2}+g'{}^{2})(v_{1}^{2}+v_{2}^{2}-v_{1}^{\prime 2}-v_{2}^{\prime 2})^{2}}{4(v_{1}^{2}+v_{2}^{2}+v_{1}^{\prime 2}+v_{2}^{\prime 2})}
\end{align}
\normalsize
\noindent
Now, if we define the angles $\tan^{2}\Tilde{\beta}=\frac{v_{1}^{2}+v_{2}^{2}}{v_{1}^{\prime2}+v_{2}^{\prime2}}$, $\tan\beta_{1}=\frac{v_{1}}{v'_{1}}$ and $\tan\beta_{2}=\frac{v_{2}}{v'_{2}}$ then the lightest scalar mass acquires the following form
\begin{align}\label{shortHiggsmass}
    m_{h1}^{2}&=m_{Z}^{2}\left(cos^{2}2\Tilde{\beta}+\frac{4}{9}\frac{g_{X}^{2}}{g^{2}+g'{}^{2}}(cos2\beta_{1}+\cos 2\beta_{2})^{2}\right) \nonumber\\
    &\approx m_{Z}^{2}\cos^{2}2\Tilde{\beta} + \Delta m_{h}^{2}
\end{align}

So in fact, when we consider the theory with additional scalar singlets and D-term, the correction term $\Delta m_{h}^{2}$ can be at the same order of tree level, and its experimental value can be explained with NMSSM and USSM. An interesting fact arises from the approximated expression for the lightest CP-even particle. Its tree level mass does not depend on the new physic's energy scale, given by $v_\chi$ and $v_\chi'$. Additionally the $\mu_{11}$ and $\mu_{22}$ factors canceled out due to the see saw mechanism, making an eigenvalue depending only on the electroweak scale VEV's, as it is expected. 

As it will be shown later and has been already mentioned, the VEV should fulfill $v_{1}^{2}+v_{1}^{\prime 2}+ v_{2}^{2}+v_{2}^{\prime 2}=246^{2}$ GeV$^2$. Then, for instance, it is found that a $125$ GeV Higgs boson can be reached for values of $g_{X}=1.06\; g $, and a large list of possible values for $v_{i}$ and $v_{i}^{\prime}$. For example considering $v_{1}=195$, $v_{2}=138$, $v'_{1}=52$, $v'_{2}=20$ and $g_{X}=0.71$ a $125$ GeV Higgs boson is found. 

\begin{figure}
    \centering
    \includegraphics[scale=0.7]{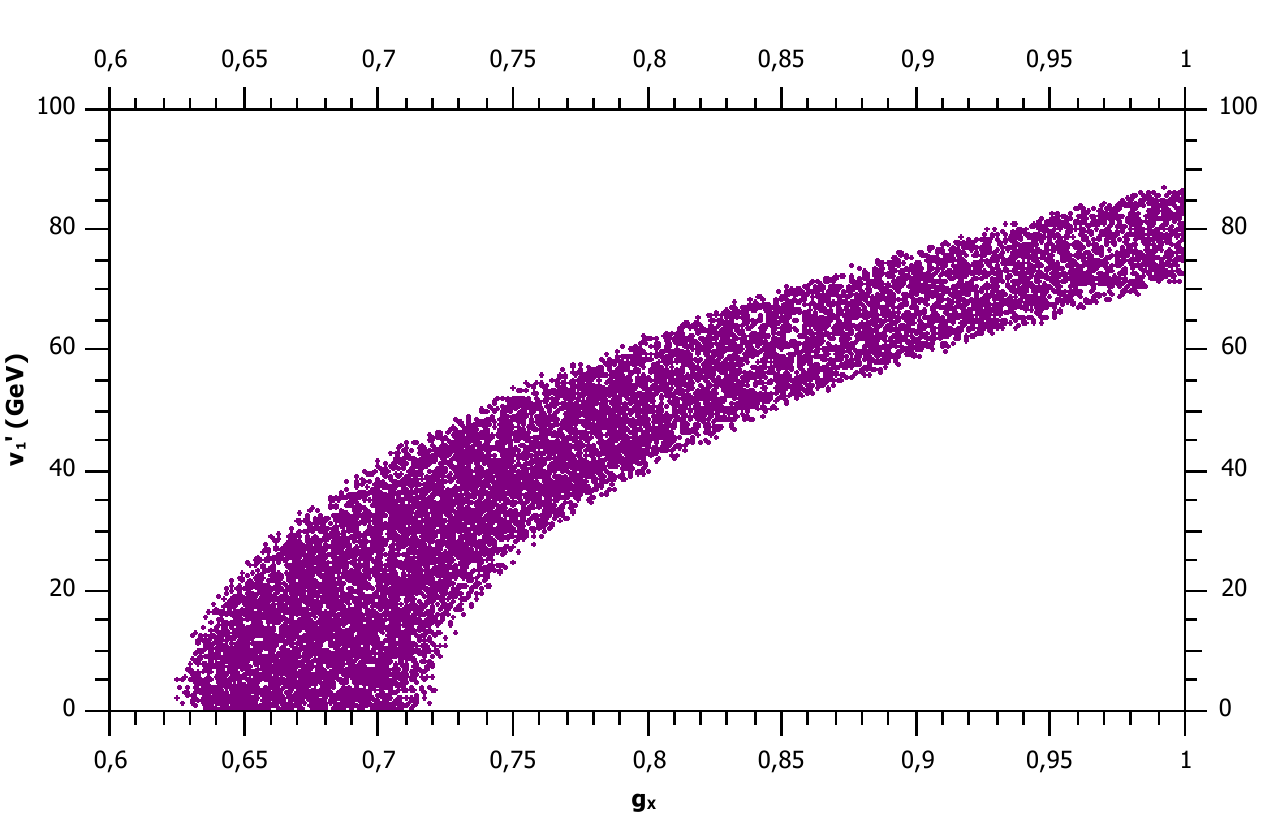}
    \caption{Region in the parameter space $v_{1}'$ vs $g_{X}$ with a Higgs mass of $125.3 \pm 0.4$ GeV at 95\% of C.L.}
    \label{fig:plotvpvsgx}
\end{figure}

In the figure (\ref{fig:plotvpvsgx}) we plot  $v'_1$ vs $g_X$ by using the expression in the eq. (\ref{Higgsobservedmass}) for the like-Higgs SM at $95\%$ of C.L. with $125.3\pm 0.4$ GeV. We take $v_1$ proportional to the top quark mass, $v_2^{\prime}$ at an intermediate value between the bottom quark and tau lepton  masses, and $v_2=\sqrt{v^2-v^2_1-v_1^{2\prime}-v^{\prime 2}_2}$. This was done for $v_{1}'$ since it is not restricted directly by the fermion mass hierarchy (FMH). The exploration in the parameter space was done by using the Montecarlo method for generating randomly the values of $v_{1}'$, $v_{2}'$ and $g_{X}$. For addressing the fermion mass hierarchy, the domains of $v_{1}$ and $v_{2}'$ were $[170 \text{GeV}, 200 \text{GeV}]$ and $[3  \text{GeV}, 7 \text{GeV}]$, respectively.  A VEV is determined by the condition \ref{VEVcondition}, thus $v_1^{\prime}=\sqrt{v^2-v^2_1-v^2_2-v_2^{\prime 2}}$. The remaining VEV could run freely, which means the its domain was $0<v_2<246$GeV.  Last by not least, the coupling parameter $g_{X}$ was explored in the domain $[0,1]$.

\begin{figure}
    \centering
    \includegraphics[scale=0.7]{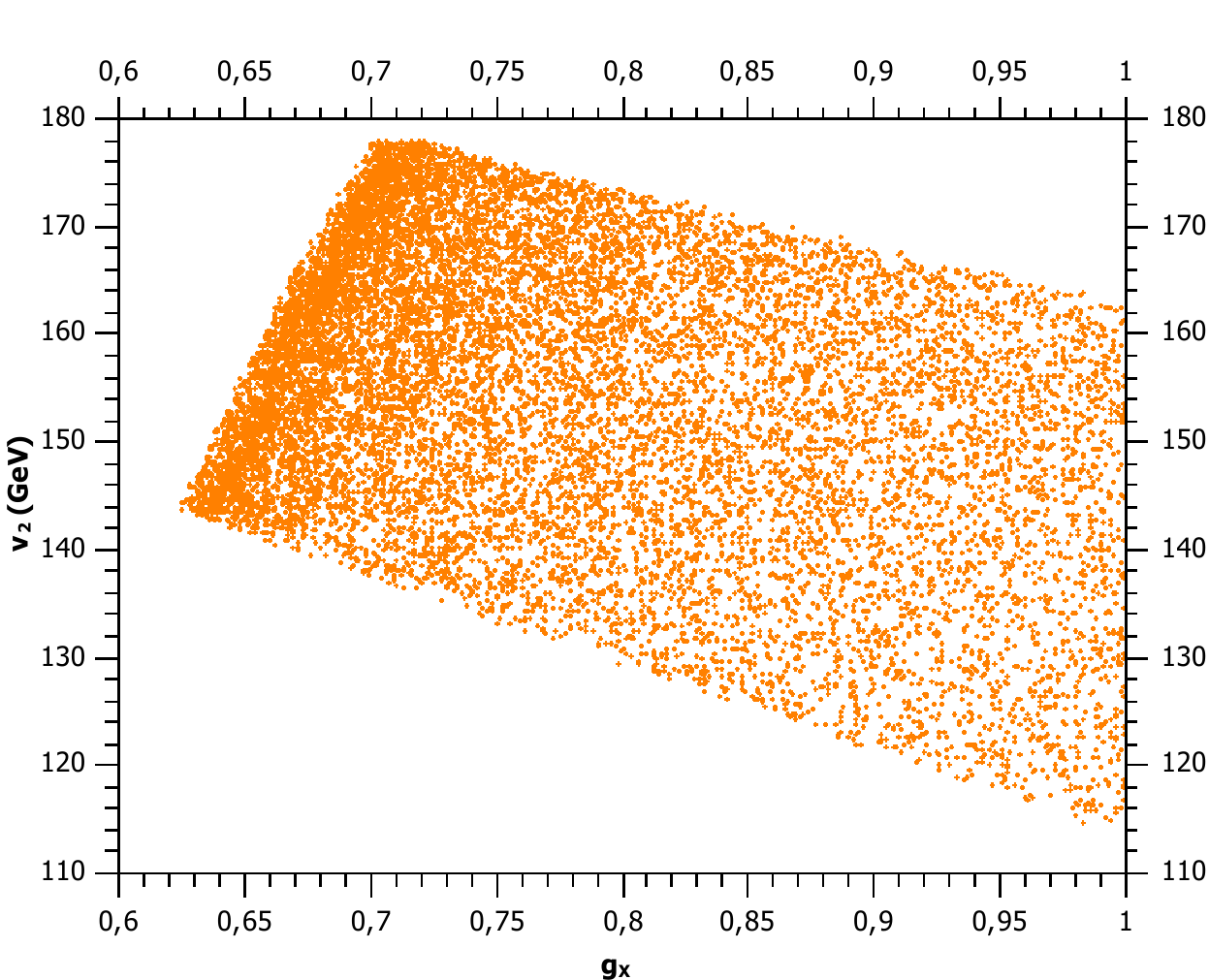}
    \caption{Region in the parameter space $v_{2}$ vs $g_{X}$ with a Higgs mass of $125.3 \pm 0.4$ GeV at 95\% of C.L.}
    \label{fig:plotv2vsgx}
\end{figure}

A similar plot is given in the figure (\ref{fig:plotv2vsgx}), where the parameter space of $v_2$ vs $g_X$ is explored within the experimental constraints at 95\% of confidence level. This is shown because $v_2$ is also not constrained by FMH.

\begin{figure}
    \centering
    \includegraphics[scale=0.7]{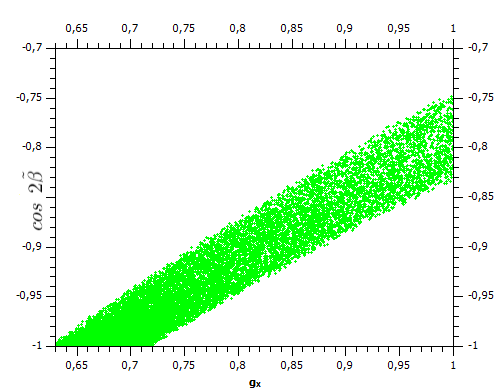}
    \caption{Region in the parameter space $\cos 2\Tilde{\beta}$ vs $g_{X}$ with a Higgs mass of $125.3 \pm 0.4$ GeV at 95\% of C.L.}
    \label{fig:cosbeta2}
\end{figure}

A clear dependence of the mass of the lightest CP-even particle with 
$$\cos 2\Tilde{\beta}=\frac{v_{1}'{}^2+v_{2}'{}^2-v_{1}{}^2-v_{2}{}^2}{v^2}$$ is found in the equation (\ref{shortHiggsmass}). In the figure (\ref{fig:cosbeta2}), the parameter space $\cos 2\Tilde{\beta}$ vs $g_{X}$ is explored. Negative values are found for $\cos 2\Tilde{\beta}$ because non-primed VEV happened to be greater than the primed ones. This behavior lies in two causes: One of them is the imposition of the FMH, where the top quark mass ($\approx v_1$) is bigger then the bottom quark mass ($\approx v'_2$) then $v_{1}> v_{2}'$. On the other hand, all along the mass expression (better seen on equation \ref{Higgsobservedmass}) there are quadratic differences between non-primed and primed VEVs. For obtaining considerable contributions to get the $125$ GeV mass, those differences have to be relatively big. Thus $v_{1},v_{2}> v_{1}',v_{2}'$ is preferred. As $g_{X}$ gets bigger, smaller differences on the VEVs are needed. Thus $|\cos 2\Tilde{\beta} |$ approaches to smaller values.

\subsection{CP-odd masses}

The mass matrix for the CP-odd particles must contain the would-be Goldstone bosons to give masses to the $Z_\mu$ and $Z'_\mu$. Such matrix in the basis $(\eta_1,\eta_1',\eta_2,\eta_2', \zeta_\chi,  \zeta_\chi',\zeta_{\sigma},\zeta_{\sigma}^{\prime})$ is given in the next equation: 
\begin{align}
   \frac{1}{2} \mathcal{M}_{\eta}&=\begin{pmatrix}
    \mathcal{M}_{\eta\eta}          & \mathcal{M}_{\eta \zeta} \\
    \mathcal{M}_{\eta\zeta}^{T}                           & \mathcal{M}_{ \zeta \zeta} 
    \end{pmatrix}.
\end{align}
$\mathcal{M}_{\eta\eta}$ contains the mixings between the CP-odd part of doublets and it can be written as
\begin{widetext}
\small
\begin{align}
    \mathcal{M}_{\eta\eta}=\begin{pmatrix}
    \frac{\mu_{11}^{2}v_{1}'}{2v_{1}}-\frac{f_{1k}v_{2}}{9v_{1}}        & \frac{\mu_{11}^{2}}{2}& \frac{f_{1k}}{9}& 0\\
   *                       & \frac{\mu_{11}^{2}v_{1}}{2v_{1}'}-\frac{f_{2k}v_{2}'}{9v_{1}'}&0&\frac{f_{2k}}{9}\\
   *&*& \frac{\mu_{22}^{2}v_{2}'}{2v_{2}}-\frac{f_{1k}v_{1}}{9v_{2}}&\frac{\mu_{22}^{2}}{2}\\
   *&*&*&\frac{\mu_{22}^{2}v_{2}}{2v_{2}'}-\frac{f_{2k}v_{1}'}{9v_{2}'}
    \end{pmatrix}.
\end{align}
\end{widetext}
\normalsize
\onecolumngrid
\noindent
The matrix accounting for the mixings between the CP-odd parts of the doublets and the singlets is $\mathcal{M}_{\eta \zeta}$, and it turns out to be:
\begin{align}
       \mathcal{M}_{\eta\chi}&= \frac{1}{9}\begin{pmatrix}
   -k_{2}v_{2}          & -k_{1}v_{2}  & \frac{1}{2\sqrt{2}}( \lambda_{2}\mu_{2}v_{2}-\tilde{\lambda}_{2}v_{2}^{\prime} ) & -\frac{1}{2\sqrt{2}}(\lambda_{1}\mu_{1}v_{2} + \lambda_{2}\mu_{\sigma}v_{2}^{\prime} )\\
    -k_{3}v_{2}'          & -k_{4}v_{2}' &  -\frac{1}{2\sqrt{2}}(\lambda_{2}\mu_{1}v_{2}^{\prime} + \lambda_{1}\mu_{\sigma}v_{2}) & \frac{1}{2\sqrt{2}}(  \lambda_{1}\mu_{2}v_{2}^{\prime}-\tilde{\lambda}_{1}v_{2}) \\
    k_{2}v_{1}          &k_{1}v_{1} &  -\frac{1}{2\sqrt{2}}(\lambda_{2}\mu_{2}v_{1} + \lambda_{1}\mu_{\sigma}v_{1}^{\prime}) & \frac{1}{2\sqrt{2}}( \lambda_{1}\mu_{1}v_{1}-\tilde{\lambda}_{1}v_{1}^{\prime}) \\
    k_{3}v_{1}'          & k_{4}v_{1}' & \frac{1}{2\sqrt{2}}( \lambda_{2}\mu_{1}v_{1}^{\prime}-\tilde{\lambda}_{2}v_{1}) & -\frac{1}{2\sqrt{2}}(\lambda_{1}\mu_{2}v_{1}^{\prime} + \lambda_{2}\mu_{\sigma}v_{1}) \\
    \end{pmatrix}. 
\end{align}

The mixings between the CP-odd part of the singlets, which are responsible of the would-be Goldstone boson due to the $U(1)_{X}$ symmetry breaking, are given by: 
\begin{align}
     \mathcal{M}_{ \zeta\zeta}=&\begin{pmatrix}
    \frac{v_{\chi}'\mu_{\chi\chi}^2}{2v_{\chi}}-\frac{k_{2}v_{1}v_{2}-k_{3}v_{1}'v_{2}'}{9v_{\chi}} & \frac{\mu_{\chi\chi}^{2}}{2} & 0 & 0\\
    * &  \frac{v_{\chi}\mu_{^{\prime 2}}^2}{2v_{\chi}'}-\frac{-k_{1}v_{1}v_{2}+k_{4}v_{1}'v_{2}'}{9v_{\chi}'} & 0 & 0 \\
    * & * &M_{\sigma}^{2} +\frac{\lambda_{2}^{2}}{4}(v_{1}^{2}+v_{2}^{\prime 2})& \frac{\mu_{\sigma\sigma}}{2} \\
    * & * & * &M_{\sigma}^{\prime 2} +\frac{\lambda_{1}^{2}}{4}(v_{2}^{2}+v_{1}^{\prime 2}) 
    \end{pmatrix}.
\end{align}

Being $M_{\sigma}^{2}$ and $M_{\sigma}^{\prime 2}$ the same coefficients defined in the scalar CP-even mass matrix. The rank of the matrix $\mathcal{M}_{\eta\eta}$ turns out to be 4, so we can be sure that there are two null eigenvalues, corresponding to the would-be Goldstone bosons needed. It is important to notice that structure of this matrix is different from $M_{h}$, but it preserves the same scale structure. That is to say, $\mathcal{M}_{\zeta\zeta} \sim \mu_{\chi\chi}, \mu_{\sigma\sigma},v_{\chi},v_{\chi}^{\prime}$ and $\mathcal{M}_{\eta\eta} \sim \mu_{ii}, \mu_{\sigma\sigma},v_{i},v_{i}^{\prime}$ which fulfill the conditions for a seesaw mechanism, $\mathcal{M}_{\zeta\zeta} >> \mathcal{M}_{\eta\eta}, \mathcal{M}_{\eta\zeta} $. When the rotation is done, the matrix reads: 
\begin{align}\label{etaSS}
    \Tilde{\mathcal{M}}_{\eta}&\approx\begin{pmatrix}
    \Tilde{\mathcal{M}}_{\eta\eta} & 0 \\
    0 & \mathcal{M}_{\zeta\zeta}
    \end{pmatrix}, \\
    \Tilde{\mathcal{M}}_{\eta\eta}&=\mathcal{M}_{\eta\eta}-\mathcal{M}_{\eta\zeta}\mathcal{M}_{\zeta\zeta}^{-1}\mathcal{M}_{\eta\zeta}^{T}.
\end{align}
\noindent
It is worth  noticing that this matrix has a big dependence on the parity breaking terms $k_{i}, i=1,2,3,4$. 

If we consider the limit in which these couplings go to zero, $k_{i} \rightarrow 0$ the mixing matrix $\mathcal{M}_{\eta\zeta}$ vanishes resulting in a $2\times 2$ isolated singlet mixing matrix $\mathcal{M}_{\zeta\zeta}(k_{i}\rightarrow 0)$ containing the $U(1)_{X}$ would-be Goldstone boson and the heaviest pseudoscalar particle predicted by the model. The two mass eigenstates are written as:
\begin{align}
  m_{\eta 5}^{2}&=0, &  m_{\eta 6}^{2}&\approx\mu_{\chi\chi}^{2}\frac{v_{\chi}^{2}+v_{\chi}'{}^{2}}{v_{\chi}v_{\chi}'}.
\end{align}
\begin{small}
\begin{align}
     m_{\eta7}^{2}=(M_{\sigma}^{2}+M_{\sigma}^{\prime 2}) +  \frac{1}{4}[\lambda_{2}^{2}(v_{1}^{2}+v_{2}^{\prime 2})+ &\lambda_{1}^{2}(v_{2}^{2}+v_{1}^{\prime 2})] \nonumber\\
     &- \sqrt{\mu_{\sigma \sigma}^{4} - \left((M_{\sigma}^{2}+M_{\sigma}^{\prime 2})+\frac{1}{4}[\lambda_{2}^{2}(v_{1}^{2}+v_{2}^{\prime 2}) - \lambda_{1}^{2}(v_{2}^{2}+v_{1}^{\prime 2})]\right)^{2}}, \nonumber\\
     m_{\eta8}^{2}=(M_{\sigma}^{2}+M_{\sigma}^{\prime 2}) +  \frac{1}{4}[\lambda_{2}^{2}(v_{1}^{2}+v_{2}^{\prime 2})+ &\lambda_{1}^{2}(v_{2}^{2}+v_{1}^{\prime 2})] \nonumber\\
     &+ \sqrt{\mu_{\sigma \sigma}^{4} - \left((M_{\sigma}^{2}+M_{\sigma}^{\prime 2})+\frac{1}{4}[\lambda_{2}^{2}(v_{1}^{2}+v_{2}^{\prime 2}) - \lambda_{1}^{2}(v_{2}^{2}+v_{1}^{\prime 2})]\right)^{2}} \nonumber
\end{align}
\end{small}

In contrast, the $4 \times 4$ matrix has a rank of 3  ensuring the existence of the would-be Goldstone boson due to $SU(2)\times U(1)$ symmetry breaking. The scheme for getting an expression for the eigenvalues is the same we used for the CP-even mass matrix. Firstly, by considering the small VEV limit we can write the matrix as:
\begin{align}
\Tilde{\mathcal{M}}_{\eta\eta}(v_{i},v_{i}'\rightarrow 0)&=\begin{pmatrix}
\frac{\mu_{11}^2}{2}\frac{v_{1}^{\prime}}{v_{1}}    & \frac{\mu_{11}^2}{2} &0&0\\
*                           &\frac{\mu_{11}^2}{2}\frac{v_{1}}{v_{1}^{\prime}}   &0&0\\
*                           &*          & \frac{\mu_{22}^2}{2}\frac{v_{2}^{\prime}}{v_{2}}&\frac{\mu_{22}^2}{2}\\
*                           &*                      &*&\frac{\mu_{22}^2}{2}\frac{v_{2}}{v_{2}^{\prime}}
\end{pmatrix} 
\end{align}
\noindent
leading to 2 heavy eigenstates which at tree level can be written as:
\begin{align}
    m_{\eta3}^{2}&\approx \mu_{11}^{2}\frac{v_{1}^{2}+v_{1}^{\prime 2}}{v_{1}v_{1}^{\prime}}, & m_{\eta4}^{2}&\approx \mu_{22}^{2}\frac{v_{2}^{2}+v_{2}^{\prime 2}}{v_{2}v_{2}^{\prime}}.
\end{align}

Finally for the lightest massive CP-odd particle it was used a small VEV limit for the general solution of the quartic equation given by Ferrari's method \cite{Ferraribib}. Having a massless state allows us to reduce the characteristic polynomial to a third grade one. Ferrari's solutions implies the cubic general solution, and through this general expressions the lighest CP-odd Higgs particles can be written as
\begin{equation}
m_{\eta 2}^{2}\approx\frac{2v^{2}(v_{1}v_{2}(k_{1}v_{\chi}^{\prime}-k_{2}v_{\chi})+v_{1}^{\prime}v_{2}^{\prime}(k_{3}v_{\chi}-k_{4}v_{\chi}^{\prime}))}{9(v_{1}^{2}+v_{1}^{\prime 2})(v_{2}^{2}+v_{2}^{\prime 2})}.
\end{equation}
The other mass corresponds to the would-be Goldstone boson associated with $Z^\mu$
\begin{equation}
m_{\eta 1}^{2}=0.
\end{equation}

\subsection{Charged scalar bosons}

In the case of the charged components of the scalar fields, the corresponding mass matrix must contain a would-be Goldstone boson that gives mass to the $W^{\mu \pm}$ gauge boson. The mass matrix in the basis $(\phi_{1}^{+},\phi_{1}^{\prime +},\phi_{2}^{+},\phi_{2}^{\prime +})$ reads
\begin{widetext}
\begin{align}
  \mathcal{M}_{C}=\begin{pmatrix} 
m^C_{11} +\frac{v_{1}'}{v_{1}}\mu_{11}^2 - \frac{1}{2}  \lambda_{2}^{2}v_{2}^{\prime 2} & \frac{g^{2}}{4}v_{1}v'_{1} + \mu_{11}^{2} & \frac{g^{2}}{4}v_{1}v_{2} +\frac{2}{9}f_{1k} & \frac{g^{2}}{4}v_{1}v'_{2} - \frac{1}{2}  \lambda_{2}^{2}v_{1}v_{2}^{\prime} \\
   * & {m}^{C\prime}_{11} +\frac{v_{1}}{v_{1}'}\mu_{11}^2 - \frac{1}{2}  \lambda_{1}^{2}v_{2}^{2}& \frac{g^{2}}{4}v'_{1}v_{2} - \frac{1}{2}  \lambda_{1}^{2}v_{2}v_{1}^{\prime}  & \frac{g^{2}}{4}v'_{1}v'_{2}+\frac{2}{9}f_{2k}  \\
   *& * & m^{C}_{22} +\frac{v_{2}'}{v_{2}}\mu_{22}^2- \frac{1}{2}\lambda_{1}^{2}v_{1}^{\prime 2} & \frac{g^{2}}{4}v_{2}v'_{2} + \mu_{22}^2 \\
   * & *& * & m^{C\prime}_{22} +\frac{v_{2}}{v_{2}'}\mu_{22}^2 - \frac{1}{2}  \lambda_{2}^{2}v_{1}^{2}
    \end{pmatrix},
\end{align}
\normalsize
where we have defined
\begin{align}
    m^C_{11}&=\frac{g^2}{4}(v_{1}'{}^2+v_{2}'{}^2-v_{2}^2)-\frac{2v_{2}}{9v_{1}}f_{1k}, &  {m}^{C\prime}_{11}&=\frac{g^2}{4}(v_{1}'{}^2+v_{2}^2-v_{2}'{}^2)-\frac{2v_{2}'}{9v_{1}'}f_{2k} , \nonumber\\
    m^{C}_{22}&=\frac{g^2}{4}(v_{1}'{}^2+v_{2}'{}^2-v_{1}^2)-\frac{2v_{1}}{9v_{2}}f_{1k}, &  m^{C\prime}_{22}&=\frac{g^2}{4}(v_{1}^2+v_{2}^2-v_{1}'{}^2)-\frac{2v_{1}'}{9v_{2}'}f_{2k} .
\end{align}
\end{widetext}
\onecolumngrid
The rank of the mass matrix for charged Higgs bosons  is 3, so there is one would-be Goldstone boson that gives masses to the $W^{\pm \mu}$ gauge boson. The procedure for obtaining the mass eigenvalues is straightforward. We perform a small VEV limit to get the heavy eigenvalues. The would-be Goldstone boson is ensured by the vanishing determinant and the lightest massive eigenvalue is found by taking a small VEV approximation in Ferrari's general solution for a cubic polynomial, giving  the following expressions:
\begin{align}
m_{G_W^\pm}^{2}&=0,\nonumber \\
m_{H_1^\pm}^{2}&\approx\frac{2v^{2}(v_{1}v_{2}(k_{1}v_{\chi}^{\prime}-k_{2}v_{\chi})+v_{1}^{\prime}v_{2}^{\prime}(k_{3}v_{\chi}-k_{4}v_{\chi}^{\prime}))}{9(v_{1}^{2}+v_{1}^{\prime 2})(v_{2}^{2}+v_{2}^{\prime 2})},\nonumber \\
m_{H_2^\pm}^{2}&\approx \mu_{11}^{2}\frac{v_{1}^{2}+v_{1}^{\prime 2}}{v_{1}v_{1}^{\prime}},\nonumber\\
m_{H_3^\pm}^{2}&\approx \mu_{22}^{2}\frac{v_{2}^{2}+v_{2}^{\prime 2}}{v_{2}v_{2}^{\prime}}.
\end{align}

\subsection{Gauge boson masses}

As consequence of the inclusion of the symmetry $U(1)_{X}$, there is a new gauge boson $B'_\mu$ which has to be included in the covariant derivate.
\begin{align}
    D_{\mu}=\partial_{\mu} -igW_{\mu}^{a}T_{a} - ig'\frac{Y}{2}B_{\mu}-ig_{X}B'_{\mu}.
\end{align}
Therefore the gauge boson masses are determined by the interaction terms, which are present in the scalar kinetic terms of the scalar fields. On one hand, the charged bosons $W^{\pm}_{\mu}=(W_{\mu}^{1}\mp W_{\mu}^{2})/\sqrt{2}$ acquire standard model like masses $M_{W}=\frac{gv}{2}$. The neutral gauge bosons $(W_{\mu}^{3},B_{\mu},B'_{\mu})$ make up a squared-mass matrix after SSB given by:
\begin{align}
    M_{0}^{2}=\frac{1}{4}\begin{pmatrix}
    g^{2} v^{2} & -gg'v^{2} & -\frac{2}{3}g g_{X} v^{2}(1+\cos^{2}\beta) \\
    * & g'{}^{2} v^{2} & \frac{2}{3}g'g_{X} v^{2}(1+\cos^{2}\beta) \\
    * & * & \frac{4}{9}g_{X}^{2} \left[V_{\chi}^{2}+(1+3\cos^{2}\beta)v^2\right]
    \end{pmatrix}, \nonumber
\end{align}
where we have defined:
\begin{align}
    v^{2}&=v_{1}^{2}+v_{2}^{2}+{v}_{1}^{\prime 2}+{v}_{2}^{\prime 2}  \label{v}\\
    \tan\beta&=\frac{\sqrt{v_{2}^{2}+{v}_{2}^{\prime 2}}}{\sqrt{v_{1}^{2}+{v}_{1}^{\prime 2}}} \equiv \frac{V_{2}}{V_{1}}\\
    V_{\chi}^{2}&\equiv v_{\chi}^{2} + {v}_{\chi}^{\prime 2}\label{vx}
\end{align}
Despite in this model we have four Higgs doublets, it recreates the same mass structure found in \cite{nosusybib} when adopting the definitions eqs. (\ref{v})-(\ref{vx}). Nevertheless, it means that the neutral boson mass eigenvalues had been already determined, and they are given by
\begin{align}
    M_{\gamma}&=0,\\
    M_{Z}&\approx\frac{gv}{2\cos\theta_{W}}, \\
    M_{Z'}&\approx \frac{g_{X}V_{\chi}}{3},\label{masadelnuevoz}
\end{align}
where $\tan\theta_{W}=\frac{g'}{g}$, as it was defined in the Standard Model. A mixing between the three neutral gauge bosons exist and it is exhibited by the following expression for the mass eigenstates:
\begin{equation}
\begin{pmatrix}
A_{\mu}\\ Z_{1\mu}\\Z_{2\mu}
\end{pmatrix}=
\begin{pmatrix}
\sin\theta_W & \cos\theta_W & 0\\
\cos\theta_W\cos\theta_Z & -\sin\theta_W\cos\theta_Z & \sin\theta_Z\\
-\cos\theta_W\sin\theta_Z & \sin\theta_W\sin\theta_Z & \cos\theta_Z\\
\end{pmatrix}
\begin{pmatrix}
W_{\mu}^{3}\\ B_{\mu}\\ B'_{\mu}
\end{pmatrix},    
\end{equation}
where $\theta_Z$ is a small mixing angle between the $Z$ and $Z'$ bosons such that in the limit $\theta_{Z} \rightarrow 0$ the standard model gauge bosons are recovered with an isolated $Z'$ boson as $Z_{1}\rightarrow Z$ and $Z_{2}\rightarrow Z'$, such mixing angle is approximately given by:
\begin{equation}\label{thetaz}
\sin\theta_Z=(1+\cos^2\beta)\frac{2g_X \cos\theta_W}{3g}\left(\frac{M_Z}{M_{Z'}}\right)^2.
\end{equation}

\subsubsection{LHC constraints on the $Z'$ mass.}
From the interaction terms of standard model fermions with the $B'_{\mu}$ gauge boson and using the mixing between the $Z$ and $Z'$ particles, one can derive the Feynman rules for fermions interacting with the masive boson $Z'_{\mu}$. Taking into account the charge quantities of particles shown in table (\ref{tab:Particle-content-A-B}), the $U(1)_X$ interaction sector of standard model fermions is shown in the following lines. For quarks we have
\begin{align}
\mathcal{L}_{int,QB'}=&\frac{g_{X}}{3}\bar{u}^1\gamma^\mu P_Lu^1 B'_{\mu} +\frac{2g_{X}}{3}\bar{u}^i\gamma^\mu P_R u^i  B'_{\mu}+\frac{g_{X}}{3}\bar{d}^1 \gamma^{\mu}P_L d^1 B'_{\mu} -\frac{g_{X}}{3}\bar{d}^i\gamma^\mu P_R d^i  B'_{\mu}\nonumber,
\end{align}
where Einstein notation convention is adopted with i=1,2,3. We also recall that $P_{L,R}=\frac{1\mp \gamma^{5}}{2}$ are the chirality projectors. In the case of the charged leptons, we have the following interaction lagrangian:
\begin{align}
\mathcal{L}_{int,eB'}=&-\frac{4g_X}{3}\bar{e}^e \gamma^\mu P_R e^e B'_{\mu}-\frac{g_X}{3}\bar{e}^\mu \gamma^\mu P_R e^\mu B'_{\mu}-g_{X}\bar{e}^\tau \gamma^\mu P_L e^\tau B'_{\mu}-\frac{4 g_X}{3}\bar{e}^\tau \gamma^\mu P_R e^\tau B'_{\mu}\nonumber 
\end{align}

 Once we have the corresponding Feynman Rules, the cross section for the $pp\rightarrow Z'\rightarrow l^{+} l^{-}$ process was calculated and it is given by:
 \begin{align}
     \frac{d\sigma}{dMdy}&=\frac{K(M)}{24\pi M^{3}}\sum_{q}P G_{q}^{\dagger}
 \end{align}
where $M=M_{ff}$ is the final state invariant mass, $y$ is the rapidity, $K(M)\approx 1.3$ resumes all leading order QED corrections and NLO QCD corrections, $P=s^{2}/[(s-M_{Z'}^{2})^{2} + M_{Z'}^{2}\Gamma_{Z'}^{2}]$ with $\sqrt{s}$ the collider CM energy together with $M_{Z'}$ and $\Gamma_{Z'}$ standing for the $Z'$ mass and total decay width respectively and $G_{q}^{\dagger}=x_{A}x_{B}[f_{q/A}(x_{A})f_{q/B}(x_{B})+ f_{q/B}(x_{B})f_{q/A}(x_{A})]$ contains the Parton Distribution Functions (PDF) $f(x)$ being $x$ the momentum fraction. In the simulation, the ratio $\Gamma_{Z'}/M_{Z'}$ was set to $0.01$ which allows us to consider the Narrow Width Approximation (NWA) for approximating the cross section to:
\begin{align}
\sigma(pp\rightarrow f\bar{f}) = \sigma(pp \rightarrow Z')BR(Z' \rightarrow f\bar{f}).
\end{align}
 
\begin{figure}[H]
    \centering
    \includegraphics[scale=0.7]{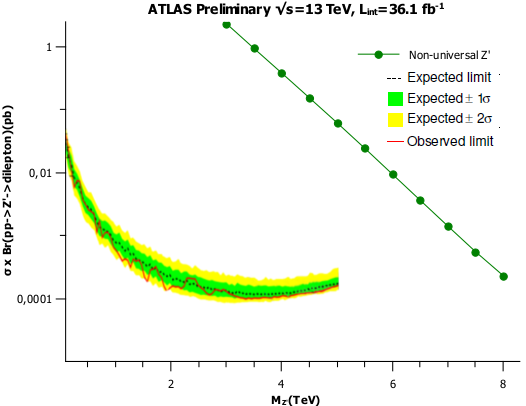}
    \caption{Observed and expected $90\%$ C.L upper limits for total cross section of dilepton production with ATLAS data. It is compared with calculated cross section for $Z'$ production times the branching ratio of the indicated decay.}
    \label{fig:ATLAScomparisondilepton}
\end{figure}

In the case of leptons, the $pp$ collision is based on a CM energy of $\sqrt{s}=14 \text{TeV}$ and a $36.1 fb^{-1}$ luminosity in agreement with ATLAS detector parameters. Furthermore, leptons have to be isolated inside a cone of angular radius $\Delta R=0.5$ in addition to a required transverse energy $E_{T}>20 \text{GeV}$ and a $|\eta|<2.5$ pseudorapidity. The results of the total cross section for the $l^{+}l^{-}$ pair production as a function of $M_{Z'}$ are shown in figure (\ref{fig:ATLAScomparisondilepton})
 
Once the $Z'$ is discovered at LHC through the DY process, it is of great importance to establish the $Z-Z'$ mixing which turns out to be model dependent. The current analysis is based on $pp$ collisions with CM energy of $\sqrt{s}=13 \text{TeV}$ collected by ATLAS ($36.1 fb^{-1}$) and CMS ($35.9 fb^{-1}$) at LHC. In particular, the process $pp\rightarrow Z' \rightarrow W^{+}W^{-}$ is considered where the coupling $Z'W^{+}W^{-}$ is only possible by the $sin\theta_{Z}$ mixing of $Z-Z'$. Consequently, by considering again the NWA the diboson production can be approximated to
\begin{align}
    \sigma(pp\rightarrow W^{+}W^{-})=\sigma(pp\rightarrow Z')BR(Z'\rightarrow W^{+}W^{-})
\end{align}
 where the expression for the partial width of $Z'\rightarrow W^{+}W^{-}$ is given by
 \begin{align}
     \Gamma_{Z'}^{\mu\nu}&=\frac{\alpha}{48}\cot^{2}\theta_{W}M_{Z'}\left(\frac{M_{Z'}}{M_{W}}\right)^{4}\left(1- 4\frac{M_{W}^{2}}{M_{Z'}^{2}}\right)^{3/2}\left[1+20\left(\frac{M_{W}}{M_{Z'}}\right)^{2} + 12\left(\frac{M_{W}}{M_{Z'}}\right)^{4}\right]\sin^{2}\theta_{Z}
 \end{align}
 with $\sin{\theta_{Z}}$ defined in eq. (\ref{thetaz}). 
 
\begin{figure}[H]
    \centering
    \includegraphics[scale=0.8]{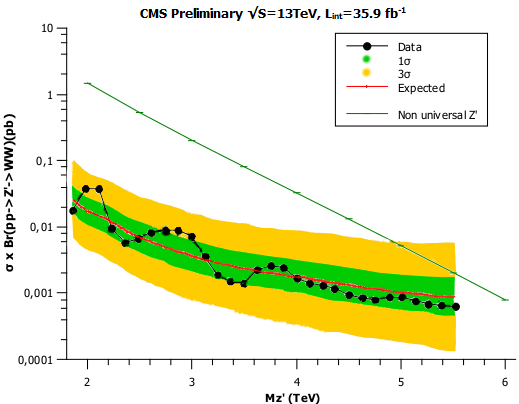}
    \caption{Observed and expected $95\%$ C.L upper limits for total cross section of diboson production with CMS data. It is compared with calculated cross section for $Z'$ production times the branching ratio of the indicated decay. }
    \label{fig:CMScomparison}
\end{figure}

\begin{figure}[H]
    \centering
    \includegraphics[scale=0.8]{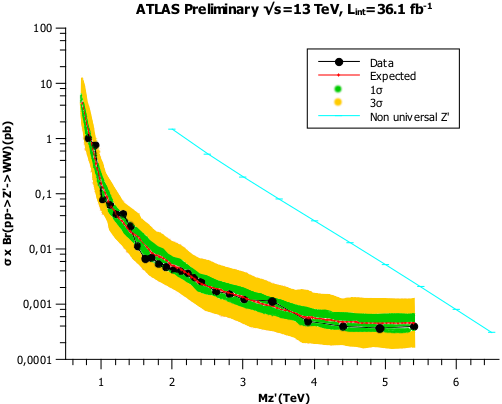}
    \caption{Observed and expected $95\%$ C.L upper limits for total cross section of diboson production with ATLAS data. It is compared with calculated cross section for $Z'$ production times the branching ratio of the indicated decay.}
    \label{fig:ATLAScomparison}
\end{figure}

 Therefore, the model was implemented in \textit{MADGRAPH5_aMC@NLO} together with \textit{PYTHIA 6} and \textit{Delphes 3} for studying the process at leading order by assuming the relevant parameters as $g_{X}=0.63$ and $\cos{\beta}=0.81$. Both of them correspond nearly to the smallest values found in the Monte Carlo simulation; specially required for the mixing angle $\theta_{Z}$ to soften experimental bounds on the $Z'$ mass. A comparison among the total cross section of the process and the available CMS \cite{CMSdataondiboson} and ATLAS \cite{ATLASdataondiboson} data on $W^{+}W^{-}$ pair production is shown in figures (\ref{fig:CMScomparison}) and (\ref{fig:ATLAScomparison}).

 The intercept between the curve obtained for the non-universal $Z'$ decaying into $W^{+}W^{-}$ and the $3\sigma$ upper limit given by CMS data shown in figure (\ref{fig:CMScomparison}), indicates a exclusion limit for a new massive boson of $M_{Z'} > 5$ TeV, in contrast with ATLAS upper limits in figure (\ref{fig:ATLAScomparison}), which provides a stronger constraint of $M_{Z'} > 5.9$ TeV


\section{Fermion sector}

\begin{table}
\caption{Fermion content of the abelian extension, non-universal $X$ quantum number and parity $\mathbb{Z}_{2}$.}
\label{tab:Particle-content-A-B}
\centering
\begin{tabular}{lll lll}\hline\hline 
\multirow[l]{3}{*}{
\begin{tabular}{l}
    Left-    \\
    Handed  \\
    Fermions
\end{tabular}
}
&\multicolumn{2}{l}{}&
\multirow[l]{3}{*}{
\begin{tabular}{l}
    Right-    \\
    Handed  \\
    Fermions
\end{tabular}
}
&\multicolumn{2}{l}{}\\ 
 &&&
 && \\ 
 &$X^{\pm}$&&
 &$X^{\pm}$&
\\ \hline\hline 
\multicolumn{6}{c}{SM Quarks}\\ \hline\hline	
\begin{tabular}{c}	
	$ \hat{q} ^{1}_{L}=\begin{pmatrix}\hat{u}^{1}	\\ \hat{d}^{1} \end{pmatrix}_{L}$  \\
	$  \hat{q} ^{2}_{L}=\begin{pmatrix}\hat{u}^{2}	\\ \hat{d}^{2} \end{pmatrix}_{L}$  \\
	$  \hat{q} ^{3}_{L}=\begin{pmatrix}\hat{u}^{3}	\\ \hat{d}^{3} \end{pmatrix}_{L}$ 
\end{tabular} &
\begin{tabular}{c}
		$\sfrac{+1}{3}^{+}$	\\
	\\	$0^{-}$	\\
	\\	$0^{+}$	\\
\end{tabular}   &
\begin{tabular}{c}
			\\
	\\		\\
	\\		\\
\end{tabular}   &
\begin{tabular}{c}
	$ \begin{matrix}\hat{u}^{1\; c }_{L}	\\ \hat{u}^{2\; c}_{L} \end{matrix}$  \\
	$ \begin{matrix}\hat{u}^{3\; c}_{L}	\\ \hat{d}^{1\; c }_{L} \end{matrix}$  \\
	$ \begin{matrix}\hat{d}^{2\; c }_{L}	\\ \hat{d}^{3\; c }_{L} \end{matrix}$ 
\end{tabular} &
\begin{tabular}{c}
	$ \begin{matrix} \sfrac{-2}{3}^{+}	\\ \sfrac{-2}{3}^{-} \end{matrix}$  \\
	$ \begin{matrix} \sfrac{-2}{3}^{+}	\\ \sfrac{+1}{3}^{-} \end{matrix}$  \\
	$ \begin{matrix} \sfrac{+1}{3}^{-}	\\ \sfrac{+1}{3}^{-} \end{matrix}$ 
\end{tabular} &
\begin{tabular}{c}
	\\
	 \\
\end{tabular}
\\ \hline\hline 

\multicolumn{6}{c}{SM Leptons}\\ \hline\hline	
\begin{tabular}{c}	
	$\hat{\ell}^{e}_{L}=\begin{pmatrix}\hat{\nu}^{e}\\ \hat{e}^{e} \end{pmatrix}_{L}$  \\
	$\hat{\ell}^{\mu}_{L}=\begin{pmatrix}\hat{\nu}^{\mu}\\ \hat{\mu}^\mu \end{pmatrix}_{L}$  \\
	$\hat{\ell}^{\tau}_{L}=\begin{pmatrix}\hat{\nu}^{\tau}\\ \hat{\tau}^{\tau} \end{pmatrix}_{L}$ 
\end{tabular} &
\begin{tabular}{c}
		$0^{+}$	\\
	\\	$0^{+}$	\\
	\\	$-1^{+}$	\\
\end{tabular} &
\begin{tabular}{c}
		\\
	\\		\\
	\\		\\
\end{tabular}   &
\begin{tabular}{c}
	$ \begin{matrix}\hat{\nu}^{e\; c}_{L}	\\ \hat{\nu}^{\mu\; c}_{L} \end{matrix}$  \\
	$ \begin{matrix}\hat{\nu}^{\tau\; c }_{L}	\\  \hat{e}^{e\; c}_{L} \end{matrix}$  \\
	$ \begin{matrix} \hat{e}^{\mu\; c }_{L}	\\  \hat{e}^{\tau\; c}_{L} \end{matrix}$ 
\end{tabular} &
\begin{tabular}{c}
	$ \begin{matrix} \sfrac{-1}{3}^{-}	\\ \sfrac{-1}{3}^{-} \end{matrix}$  \\
	$ \begin{matrix} \sfrac{-1}{3}^{-}	\\ \sfrac{+4}{3}^{-} \end{matrix}$  \\
	$ \begin{matrix} \sfrac{+1}{3}^{-}	\\ \sfrac{+4}{3}^{-} \end{matrix}$ 
\end{tabular} &
\begin{tabular}{c}
	  \\
	 \\
\end{tabular}
\\ \hline\hline 

\multicolumn{6}{c}{Non-SM Quarks}\\ \hline\hline	
\begin{tabular}{c}	
	$\hat{\mathcal{T}}_{L}$	\\		\\
	$\mathcal{J}_{L}^{1}$	\\	$\mathcal{J}_{L}^{2}$	
\end{tabular} &
\begin{tabular}{c}
	$\sfrac{+1}{3}^{-} $	\\	\\	
	$ 0^{+} $           	\\	$ 0^{+} $
\end{tabular}   &
\begin{tabular}{c}
		\\		\\	
		\\		
\end{tabular}   &

\begin{tabular}{c}
	$\hat{\mathcal{T}}_{L}^{c}$	\\		\\
	$\hat{\mathcal{J}}_{L}^{c}$	\\	$\hat{\mathcal{J}}_{L}^{c \ 2}$	
\end{tabular} &
\begin{tabular}{c}
	$\sfrac{-2}{3}^{-} $	\\	              	\\	
	$\sfrac{-1}{3}^{+} $	\\	$\sfrac{-1}{3}^{+} $
\end{tabular} &
\begin{tabular}{c}
		\\		\\	
		\\		
\end{tabular}
\\ \hline\hline

\multicolumn{6}{c}{Non-SM Leptons}\\ \hline\hline	
\begin{tabular}{c}	
    $\hat{E}_{L}$	\\
    $\hat{\mathcal{E}}_{L}$	\\
\end{tabular} &
\begin{tabular}{c}
	$-1^{+}$	    	\\
	$\sfrac{-2}{3}^{+}$	\\
\end{tabular} &
\begin{tabular}{c}
			\\
			\\
		
\end{tabular}   &
\begin{tabular}{c}	
    $\hat{E}_{L}^{c}$	\\
    $\hat{\mathcal{E}}_{L}^{c}$	\\
\end{tabular} &
\begin{tabular}{c}
	$\sfrac{+2}{3}^{+}$		\\
	$+1^{+}$	            \\
\end{tabular} &
\begin{tabular}{c}
			\\
			\\
	
\end{tabular}
\\	\hline\hline 

\multicolumn{3}{c}{Majorana Fermions} & 
\begin{tabular}{c}	
	$\mathcal{N}_{R}^{1,2,3}$	\\
\end{tabular} &
\begin{tabular}{c}
	$0^{-}$	\\
\end{tabular} &
\begin{tabular}{c}
	\\
		
\end{tabular}
\\	\hline\hline 
\end{tabular}
\end{table}
\subsection{Quark Sector}
According to the $SU(2)_{L}\otimes U(1)_{Y}\otimes U(1)_{X}\otimes Z_{2}$ symmetry, the most general Yukawa superpotencial for the quark superfields is given by:\\
\begin{align}
    W_{Q}&=\hat{q}_{L}^{1}\hat{\Phi}_{2}h_{2u}^{12}\hat{u}_{L}^{2\; c} + \hat{q}_{L}^{2}\hat{\Phi}_{1}h_{1u}^{22}\hat{u}_{L}^{2\; c} + \hat{q}_{L}^{3}\hat{\Phi}_{1}h_{1u}^{3k}\hat{u}_{L}^{k\; c} - \hat{q}_{L}^{3}\hat{\Phi '}_{2}h_{2d}^{3j}\hat{d}_{L}^{j\; c}+ \hat{q}_{L}^{1}\hat{\Phi}_{2}h_{2T}^{1}\hat{\mathcal T}_{L}^{c} +\hat{q}_{L}^{2}\hat{\Phi}_{1}h_{1T}^{2}\hat{\mathcal T}_{L}^c\nonumber\\
    &-\hat{q}_{L}^{1}\hat{\Phi '}_{1}{h}_{1J}^{1a}\hat{\mathcal J}_{L}^{a\; c} - \hat{q}_{L}^{2}\hat{\Phi '}_{2}{h}_{2J}^{2a}\hat{\mathcal J}_{L}^{a\; c} + \hat{\mathcal T}_{L}\hat{\chi}'h_{\chi'}^{ T}\hat{\mathcal T}_{L}^{c} - \hat{\mathcal J}_{L}^{a}\hat{\chi}h_{\chi}^{J ab}\hat{\mathcal J}_{L}^{b\; c} +\hat{\mathcal T}_{L}\hat{\chi '}h_{\chi' u}^2\hat{u}_{L}^{2\; c} + \hat{\mathcal{J}}_{L}^{a}\hat{\sigma} h_{\sigma}^{J aj}\hat{d}_{L}^{j c } \nonumber\\
    &+ \hat{\mathcal{T}}_{L}\hat{\sigma}^{\prime}h_{\sigma^{\prime}}^{Tk}\hat{u}_{L}^{k c}
\end{align}
where $j=1,2,3$ labels the down type singlet quarks, $k=1,3$ labels the first and third quark doublets, and $a=1,2$ is the index of the exotic $\mathcal{J}_{L}^{a}$ and $\mathcal{J}_{L}^{c a}$ quarks. It can be seen that this general superpotencial match the non supersymmetrical one given in ref \cite{nosusybib} if we promote the conjugate Higgs fields $\Tilde{\phi}_{i}=i\sigma_{2}\phi_{i}^{*}$ to the independent ones $\Phi '_{i}$ required for a suitable anomaly cancellation. As a consequence, taking the VEV of all scalar fields, the quark mass matrices at tree level have an identical structure to its non-SUSY counterpart, as it can be seen in the following equations. For the up quark sector has 
\begin{equation}
\mathcal{M_{U}}=
\begin{pmatrix}
M_U & M_{UT} \\
M_{TU} & M_T
\end{pmatrix}
\end{equation}
where
\begin{align}
    M_{U}&=\frac{1}{\sqrt{2}}\begin{pmatrix}
   0 & h_{2u}^{12}v_{2} & 0 \\
    0 & h_{1u}^{22}v_{1} & 0 \\
    h_{1u}^{31}v_{1} & 0 &h_{1u}^{33}v_{1} 
    \end{pmatrix}, \label{QMi}  \;\;
    M_{UT}=\frac{1}{\sqrt{2}}\begin{pmatrix}
    h_{2T}^{1}v_{2} \\
    h_{1T}^{2}v_{1} \\
    0
    \end{pmatrix},
\end{align}
\begin{align}
    M_{TU}&=\frac{v'_{\chi}}{\sqrt{2}}\begin{pmatrix}
    0 & h_{\chi' u}^2 & 0
    \end{pmatrix} ,\;\;\;\;\;\;\;\;\;\;\;\;
    M_{T}=\frac{v'_{\chi}}{\sqrt{2}}g_{\chi' T}, \label{QMf} 
\end{align}
and for the down quark, the matrices can be written as
\begin{equation}
\mathcal{M_D}=
\begin{pmatrix}
M_D & M_{DJ} \\
M_{JD} & M_J
\end{pmatrix}
\end{equation}
where
\begin{align}
M_{D}&=\frac{v'_{2}}{\sqrt{2}}\begin{pmatrix}
    0 & 0 & 0 \\
    0 & 0 & 0 \\
    h_{2d}^{31} & h_{2d}^{32} & h_{2d}^{33} 
    \end{pmatrix},\;\;
    M_{DJ}=\frac{1}{\sqrt{2}}\begin{pmatrix}
    {h}_{1J}^{11}v'_{1} & {h}_{1J}^{12}v'_{1} \\
    {h}_{2J}^{21}v'_{2} &{h}_{2J}^{22}v'_{2} \\
    0 & 0 
    \end{pmatrix}, \;\;\nonumber\\
     M_{JD}&=\begin{pmatrix}
    0 & 0 & 0 
    \end{pmatrix},\;\;\;\;\;\;\;\;\;\;\;\;
    M_{J}=\frac{v_{\chi}}{\sqrt{2}}\begin{pmatrix}
    g_{\chi J}^{11} & g_{\chi J}^{12} \\
    g_{\chi J}^{21} & g_{\chi J}^{22}
    \end{pmatrix}. \label{QMfd}
\end{align}
It is worth to notice that up-like quarks acquire mass from $\Phi_{i}$ Higgs fields while the down-like quarks do it from the $\Phi '_{i}$ ones. Since the matrix structure is identical to its non-SUSY counterpart, the same analysis and eigenvalues gotten in \cite{nosusybib} can be done, taking care now that the down-like eigenvalues are coupled to primed Higgs VEV. The mass expressions can be approximated to:
 \begin{align}
m_{u}^{2}&=0, \;\;\;\;\;\;\;\;\;\;\;\;\;\;\;\;\;\;\;\;\;\;\;\;\;\;\;\;\;\;\;\;\;\;\;\;\;\;\;\;\;
m_{c}^{2}=\frac{1}{2}v_1^2\frac{\left[h_{1u}^{22}g_{\chi' T}-h_{1T}^{2}h_{\chi' u}^2\right]^2}{(g_{\chi' T})^2+(h_{\chi' u}^2)^2},\nonumber \\
m_{t}^{2}&=\frac{1}{2}v_1^2\left[(h_{1u}^{31})^2+(h_{1u}^{33})^2\right],\;\;\;\;\;\;\;\;\;\; m_{T}^{2}=\frac{1}{2}v_\chi'{}^2\left[(g_{\chi' T})^2+(h_{\chi' u}^2)^2\right].
 \end{align}
The hierarchy between top and charm masses comes from the see-saw rotation with the heavy $\mathcal{T}$ quark, which can be observed from the Yukawa coupling differences for the charm quark mass. From the Charm and Top quarks physical masses it is known that $\frac{m_{c}}{m_{t}}\approx 7\times  10^{-3}$ which in terms of Yukawa couplings reads:
\begin{align}
    7\times 10^{-3} \approx \frac{h_{1u}^{22}g_{\chi' T}-h_{1T}^{2}h_{\chi' u}^2}{\sqrt{(g_{\chi' T})^2+(h_{\chi' u}^2)^2}\sqrt{(h_{1u}^{31})^2+(h_{1u}^{33})^2}}
\end{align}
To get an estimate of the above expression considering the mass hierarchy we can consider Yukawa couplings $(g_{\chi' T})^{2}$, $(h_{\chi' u}^2)^{2}$, $(h_{1u}^{31})^2$ and $(h_{1u}^{33})^2$ of order 1 which implies a phenomenologically viable relation among Yukawas:

\begin{align}
    1.4\times 10^{-2} \approx h_{1u}^{22}g_{\chi' T}-h_{1T}^{2}h_{\chi' u}^2
\end{align}

For the the down-quarks, the mass expressions can be approximated to:
\begin{align}m_{d}^2&=0,\;\;\;\;\;\;\;\;\;\;\;\;\;\;\;\;\;\;\;\;\;\;\;\;\;\;
m_{s}^{2}=0,\;\;\;\;\;\;\;\;\;\;\;\;
m_{b}^{2}=\frac{1}{2}v_2'{}^2\left[(h_{2d}^{31})^2+(h_{2d}^{32})^2+(h_{2d}^{33})^2\right],\nonumber \\
m_{J^1}^{2}&=\frac{1}{2}v_\chi^2(g_{\chi J}^{11})^2,\;\;\;\;\;\;\;\;\;\;\;\;
m_{J^2}^{2}=\frac{1}{2}v_\chi^2(g_{\chi J}^{22})^2.
 \end{align}

The $m_{u}^{2}$, $m_{d}^{2}$ and $m_{s}^{2}$ masses are equal to zero but they can be obtained by radiative corrections taking into account the SUSY contribution due to the respective superpartners as we will show later. According to the above expressions, it is the scalar particle $\phi_{1}$ which gives mass to the top and charm quarks through $v_{1}$ and similarly $v_{2}^{\prime}$ provides a mass value for the bottom quark. However the mass difference between charm and top quarks is fully dependent on the Yukawa couplings values which allows us to assume $v_{1} \approx m_{t}$ and $v_{2}^{\prime}\approx m_{b}$. Looking at the exotic sector, they are governed entirely by the values of $v_{\chi}$ and $v_{\chi}^{\prime}$ whose values are expected to be at least in the TeV scale. Furthermore, it agrees with recent experimental results which exclude exotic quarks with masses bellow $800 \textit{GeV}$ \cite{dataonexotic}

\subsection{Lepton sector}

Analogously, the lepton superpotencial corresponds to the non-SUSY Yukawa lagrangian; the fields are promoted to superfields and the conjugate Higgs fields promoted to the primed ones. Then the superpotenial reads as follows
\begin{align}
    W_{L}&= \hat{\ell}_{L}^{p}\hat{\Phi_{2}}h_{2\nu}^{pq}\hat{\nu}_{L}^{q\; c} -\hat{\ell}_{L}^{p}\hat{\Phi '_{2}}{h}_{2e}^{p\mu}\hat{e}_{L}^{\mu\; c}
    - \hat{\ell}_{L}^{\tau}\hat{\Phi '_{2}}{h}_{2e}^{\tau r}\hat{e}_{L}^{r\; c} 
    - \hat{\ell}_{L}^{p}\hat{\Phi '_{1}}{h}_{1E}^{p}\hat{E}_{L}^{c}+ \hat{E}_{L}\hat{\chi '}{g}_{\chi' E}\hat{E}_{L}^{c}  \nonumber \\
    &- \hat{E}_{L}\mu_{E}\hat{\mathcal{E}}_{L}^{c} + \hat{\mathcal{E}}_{L}\hat{\chi}g_{\chi\mathcal{E}}\hat{\mathcal{E}}_{L}^{c}  - \hat{\mathcal{E}}_L\mu_{\mathcal{E}}\hat{E}_{L}^{c} +\hat{\nu}_{L}^{j\; c}\hat{\chi} ' {h}_{\chi}^{\prime N\; ij}\hat{N}_{L}^{i\; c}
        + \frac{1}{2}\hat{N}_{L}^{i\; c} M_{ij}\hat{N}_{L}^{j\; c} \nonumber\\
        &+ \hat{E}_{L}\hat{\sigma} h_{\sigma}^{e^{c} p}\hat{e}_{L}^{c r} + \hat{\mathcal{E}}_{L}\hat{\sigma}^{\prime}h_{\sigma'}^{e^{c} \mu}\hat{e}_{L}^{\mu c},\label{WL}
\end{align}
where $p=e,\mu$ , $q=e,\mu,\tau$, $r=e,\tau$ and $i,j$ label the right handed and Majorana neutrinos. The superpotential presented in the equation (\ref{WL}) generates the same mass matrix structure as well for the neutral and charged leptons when the VEV is taken by the fields, compared to the non-SUSY model.
 
\subsubsection{Charged leptons masses at tree level}

The mass matrix for the charged leptons follows also the same structure as the one obtained in the non-supersymmetrical model. It is shown right ahead:
\begin{align}
    \mathcal{M}_{E}&=\frac{1}{\sqrt{2}}\left(\begin{array}{ c c c |c c}\\
    0                           & h_{2e}^{e\mu}v'_{2}     & 0 &  h_{1e}^{E}v'_{1}    & 0 \\
    0                           & h_{2e}^{\mu\mu}v'_{2}   & 0 &  h_{1\mu}^{E}v'_{1}  & 0 \\
    h_{2e}^{\tau e}v'_{2}  & 0                            & h_{2e}^{\tau\tau}v'_{2} & 0 & 0 \\ \hline
    0 & 0 & 0 & {g}_{\chi' E}v'_{\chi} & -\mu_{E} \\
    0 & 0 & 0 & -\mu_{\mathcal{E}} & g_{\chi\mathcal{E}}v_{\chi}  \\
    \end{array} \right)
\end{align}
Just as it happened in the old model, the electron remains massless at tree level. Therefore radiative corrections must be employed to explain the mass feature of such particle. The mass eigenvalues at leading order are given by:

\begin{align}
m_{e}^{2}&=0, \;\;\;\;\;\;\;\;\;\;\;\;\;\;\;\;\;\;
m_{\mu}^{2}=\frac{1}{2}v_2'{}^{2}\left[(h_{2e}^{e\mu})^2+(h_{2e}^{\mu\mu})^2\right], \;\;\;
m_{\tau}^{2}=\frac{1}{2}v_2'{}^2\left[(h_{2e}^{\tau e})^2+(h_{2e}^{\tau \tau})^2\right]\nonumber\\
m_{E}^{2}&=\frac{1}{2}g_{\chi' E}^2\; v_\chi'{}^2, \;\;\;\;\;\;
m_{\mathcal{E}}^{2}=\frac{1}{2}g_{\chi \mathcal{E}}^2\; v_\chi^2
 \end{align}

Despite $V_{2}^{\prime}$ also gives mass to the bottom quark, it is a good result that it also gives mass to the charged leptons since they are of the same scale order. So, in agreement with the particle physical mass values the ratio between $\mu$ and $\tau$ lepton masses is approximately $0.14$, leaving us with the following relation among Yukawa Couplings:
\begin{align}
    0.14 \approx \frac{\sqrt{(h_{2e}^{e\mu})^2+(h_{2e}^{\mu\mu})^2}}{\sqrt{(h_{2e}^{\tau e})^2+(h_{2e}^{\tau \tau})^2}}
\end{align}

\subsubsection{Neutrino masses at tree level}

As for neutrinos, the mass matrix involves the Dirac and Majorana terms in order to provide a mass spectrum via the inverse seesaw mechanism (ISS). In the basis $(\nu_{L}^q,\nu_{L}^q{}^{C},N_{L}^{i}{}^C)$, such matrix reads:
\begin{align}
\mathcal{M}_{\nu} &=\begin{pmatrix}
    0 & m_{D}^{T} & 0 \\
    m_{D} & 0 & M_{D}^{T} \\
    0 & M_{D} & M_{M}
    \end{pmatrix},     
\end{align}
where the block matrices constituting the neutrino mass matrix are given by:
\begin{align}
 &m_{D}=\frac{v_{2}}{\sqrt{2}}\begin{pmatrix}
    h_{2\nu}^{ee} & h_{2\nu}^{e\mu} & h_{2\nu}^{e\tau} \\
    h_{2\nu}^{e\mu} & h_{2\nu}^{\mu\mu} & h_{2\nu}^{\mu\tau} \\
    0 & 0 & 0
    \end{pmatrix},\ \ \
     (M_{D})^{ij}=\frac{v'_{\chi}}{\sqrt{2}}({h}_{\chi}^{\prime \nu})^{ij}, \ \ \ \ \  (M_{M})_{ij}=\frac{1}{2}M_{ij}.
\end{align}

For the ISS to work, the assumption on small Majorana coupling constants is made, $m_{D}\gg M_{M}$. Therefore, it can be shown that the matrix $\mathbb{M}_{\nu}$ can be approximately block diagonalized:
\begin{align}
\mathbb{V}_{SS}^{T}\mathcal{M}_{\nu}\mathbb{V}_{SS}\approx\begin{pmatrix}
m_{light}&0\\
0&m_{heavy}
\end{pmatrix},
\end{align}
where $m_{light}=m_{D}^{T}(M_{D})^{-1}M_{M}(M_{D})^{-1}m_{D}$ is the $3\times3$ mass matrix for the light neutrinos and it must explain the observed mixing parameters in the PMNS matrix. The rotation matrix $\mathbb{V}_{S}$ can be calculated by:
\begin{equation}
\mathbb{V}_{SS}=\begin{pmatrix}
I&\Theta_{\nu}\\
-\Theta_{\nu}^{T}&I
\end{pmatrix} , \;\;\;
\Theta_{\nu}=\begin{pmatrix}
0&M_{D}^{T}\\
M_{D}&M_{M}
\end{pmatrix}^{-1}\begin{pmatrix}
m_{D}\\0
\end{pmatrix}.
\end{equation}
Lastly, the $m_{heavy}$ matrix involves the mixings of the exotic neutral leptons, and it is given by:
\begin{align}
m_{heavy}\approx\begin{pmatrix}0&M_{D}^{T}\\
M_{D}&M_{M}
\end{pmatrix}  .  
\end{align}

Since the same structure as the non-SUSY model is followed also by the sector of neutral leptons, the same constraints are applied for the Yukawa parameters for explaining the quadratic difference of masses $\Delta m^{2}$ and mixing angles in the neutrino oscillation \cite{exposcilacion}. The parameter values are then shown in \cite{nosusybib}.

\subsection{Fermion masses at one loop level}

\begin{figure}
    \centering
    \includegraphics[scale=0.8]{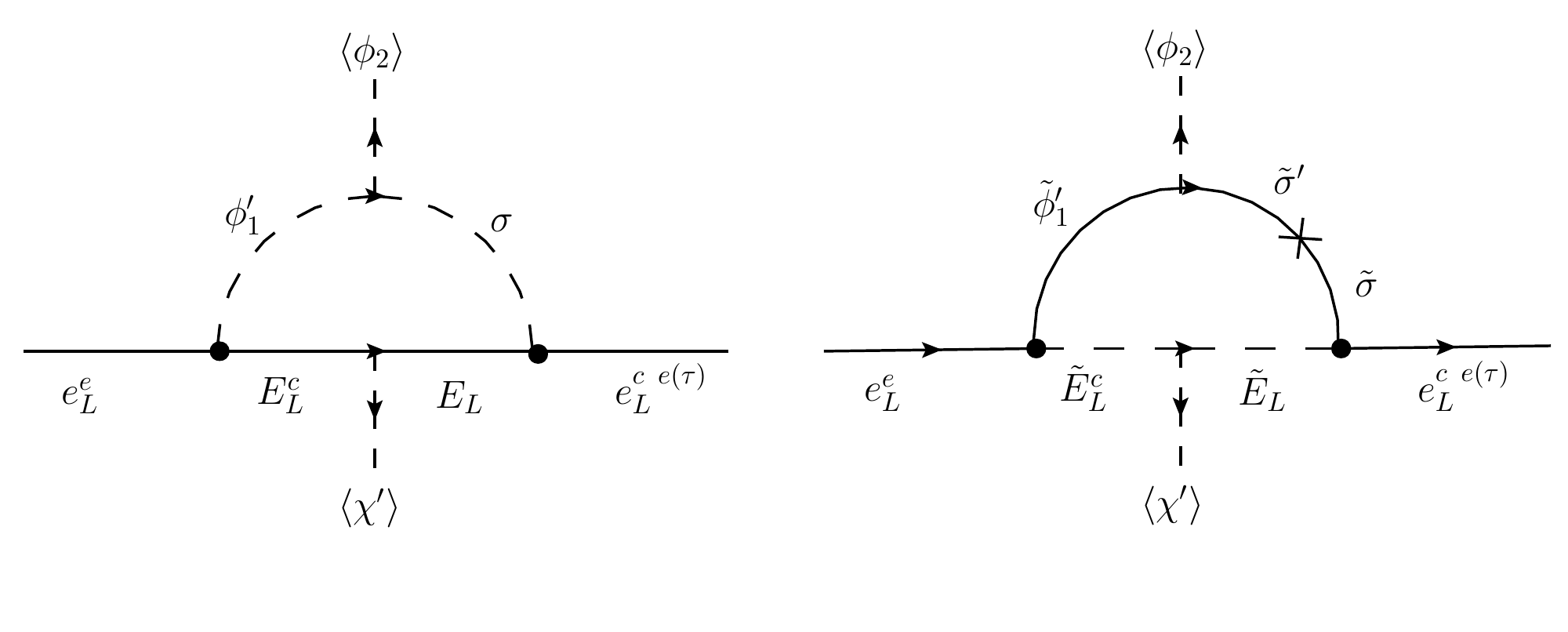}
    \caption{One loop corrections to the leptons due to exotic fermions, sfermions and Higgsinos.}
    \label{fig:1-loopforleptons}
\end{figure}

\begin{figure}[H]
    \centering
    \includegraphics[scale=0.8]{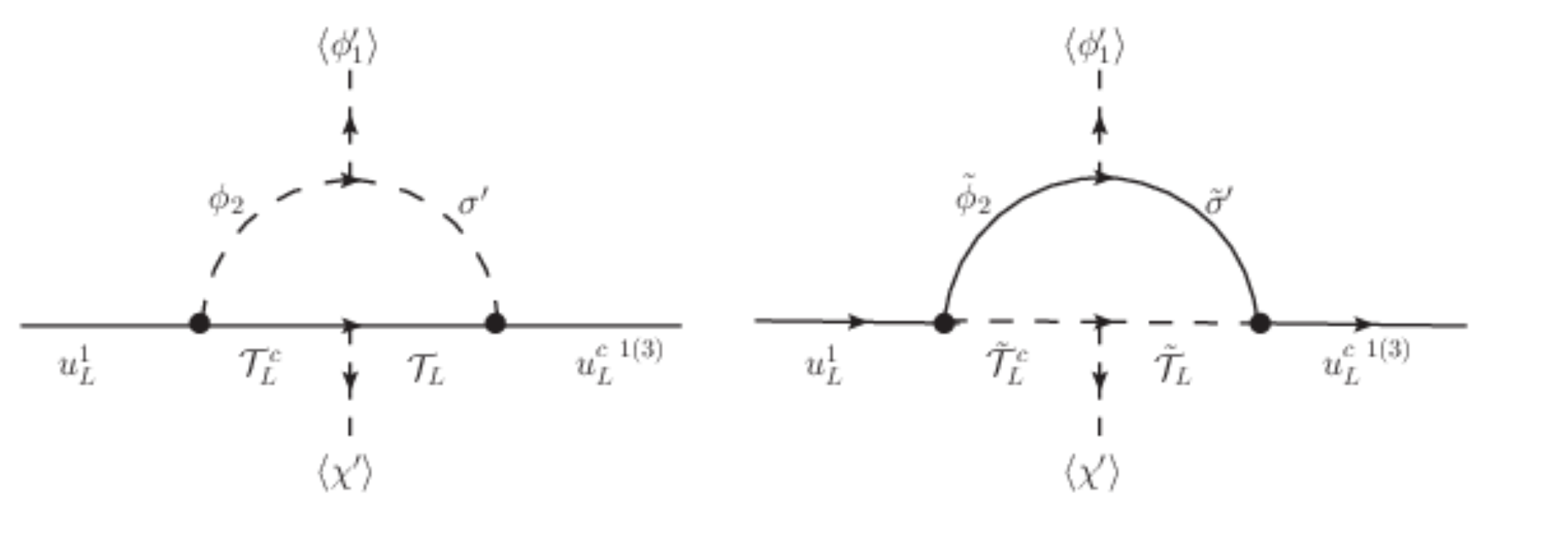}
    \caption{One loop corrections to the quark up  due to scalar singlets, exotic quarks, squarks and Higgsinos.}
    \label{fig:1-loopforquarksup}
\end{figure}

\begin{figure}
    \centering
    \includegraphics[scale=0.8]{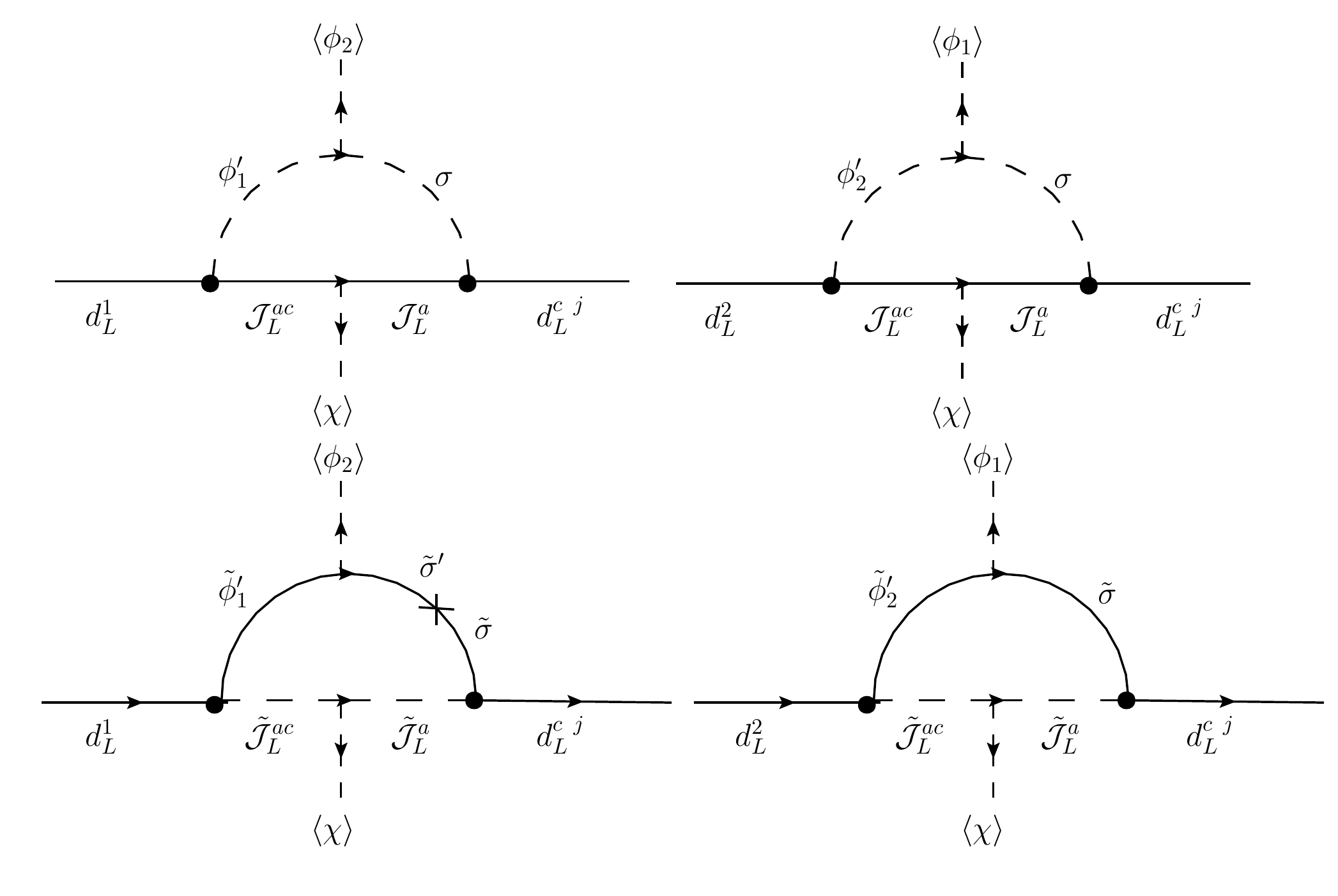}
    \caption{One loop corrections to the quarks down and strange due to scalar singlets, exotic quarks, squarks and Higgsinos.}
    \label{fig:1-loopforquarksdown}
\end{figure}

As it was seen in the previous section, the electron and the up, down and strange quarks turned out to be massless at tree level. However, since the physical mass of these particles is considerably small in comparison with the other particles and the model energy scale, they are expected to acquire a finite mass value through radiative corrections. In fact, it is performed via $\hat{\sigma}$ and $\hat{\sigma}^{\prime}$ superfields couplings, as it is shown in figures \ref{fig:1-loopforleptons}, \ref{fig:1-loopforquarksdown} and \ref{fig:1-loopforquarksup}, which allows the transition between SM fermions and the exotic ones resulting in a non-zero value for their masses . 

Considering first the electron mass, the vertices are generated by the following parts of the superpotential and the soft-breaking potential terms:

\begin{align}
    W_{\phi}&\Rightarrow\lambda_{1}\Phi_{1}^{\prime}\Phi_{2}\sigma^{\prime} -\mu_{\sigma}\hat{\sigma}^{\prime}\hat{\sigma}&
    W_{L}&\Rightarrow\hat{E}_{L}\hat{\sigma} h_{\sigma}^{e^{c} r}\hat{e}_{L}^{r c} + \hat{\ell}_{L}^{e}\hat{\Phi '_{1}}{h}_{1E}^{e}\hat{E}_{L}^{c}, 
\end{align}
\noindent
where $r=e,\tau$ and the couplings $\lambda_{1}$, $h_{\sigma}^{e^{c} k}$ and $h_{1E}^{e}$ are dimensionless Yukawa coupling constants, but $\tilde{\lambda}_{1}$ and $\mu_{\sigma}$ are mass unit parameters from the scalar potential. Considering the figure \ref{fig:1-loopforleptons}, the non-SUSY contributions are given by:

\begin{align}
    v_{2}\Sigma _{11(13)}^{NS}=\frac{-1}{16\pi ^2}\frac{v_{2}}{\sqrt{2}}\frac{\lambda_{1}\mu_{\sigma}h_{\sigma}^{ec e(\tau)}h_{1E}^{e}}{M_{E}}C_0\left(\frac{m_{h1}^{\prime}}{M_{E}},\frac{m_{\sigma}^{\prime}}{M_E}\right).
\end{align}
\noindent
with
\begin{align}\label{C0}
    C_{0}(\hat{m}_{1},\hat{m}_{2}) &= \frac{1}{(1-\hat{m}_{1}^{2})(1-\hat{m}_{2}^{2})(m_{1}^{2}-\hat{m}_{2}^{2})}\left[\hat{m}_{1}^{2}\hat{m}_{2}^{2}Ln\left(\frac{\hat{m}_{1}^{2}}{\hat{m}_{2}^{2}}\right) + \hat{m}_{2}^{2}Ln(\hat{m}_{2}^{2})- \hat{m}_{1}^{2}Ln(\hat{m}_{1}^{2})\right],
\end{align}
\noindent
where $M_{E}$ is the exotic charged fermion mass, $m_{h1}^{\prime}$ is the corresponding neutral scalar field mass related with the $\hat{\Phi}_{1}^{\prime}$ superfield, $m_{\sigma}^{\prime}$ is the scalar field mass corresponding to the $\hat{\sigma}$  superfield and  $C_{0}$ is the Veltmann-Passarino function evaluated for $p^{2}=0$ given by eq. (\ref{C0}). In the radiative correction calculation, a rotation to the mass eigenstate basis for the exotic fermion was not done by assuming a mixing angle which suppresses the mixing among them, even though, in the SUSY contribution the rotation matrix is written explicitly because those zero tree level mass terms prevent the implementation of a renormalization scheme. It implies the presence of divergences in the $B_{0}$ Veltmann-Passarino function which are cancelled out by the Super GIM mechanism resulting then in finite mass terms and a renormalizable theory. Therefore, the supersymmetric contributions to the radiative correction are given by:
\begin{align}
    v_{2}\Sigma_{11(13)}^{S}(p^{2}=0) = -\frac{1}{32\pi^{2}}&\frac{v_{2}}{\sqrt{2}}\sum_{n=1}^{10}\sum_{k=1}^{2}Z_{L}^{\tilde{E}_{L}^{c}n}Z_{L}^{\tilde{E}_{L}n} Z_{L}^{\sigma k}Z_{L}^{\sigma^{\prime} k} \lambda_{1}\mu_{\sigma}h_{\sigma}^{ec e(\tau)}h_{1E}^{e}\times \\
    &\times\left[\frac{(\tilde{m}_{\sigma  k}+\tilde{m}_{h_{1}}^{\prime})^{2}}{\tilde{M}_{L_{n}}^{2}}C_{0}\left(\frac{\tilde{m}_{h1}^{\prime}}{\tilde{M}_{L_{n}}},\frac{\tilde{m}_{\sigma  k}}{\tilde{M}_{L_{n}}} \right) + \tilde{m}_{h1}^{\prime 2}B_{0}(0,\tilde{m}_{\sigma}^{\prime},\tilde{M}_{L_{n}}) + \tilde{m}_{\sigma k}^{2}B_{0}(0,\tilde{m}_{h1}^{\prime},\tilde{M}_{L_{n}}) \right] \nonumber
\end{align}
\noindent
where $\tilde{L}_{n}$ are the charged left sleptons mass eigenstates, $Z_{L}^{\sigma (\sigma') k}$ is the rotation matrix that connects $\sigma$ ($\sigma'$)  with its mass eigenstates with eigenvalues $m_{\sigma k}$ which are running inside the loop. $Z_{L}^{\tilde{E}_{L}^{c}n}$ and $Z_{L}^{\tilde{E}_{L}n}$ are the rotation matrices which connect the exotic sleptons $\tilde{E}_{L}$, $\tilde{E}_{L}^{c}$ with the slepton mass eigenstates $\tilde{L}_{n}$ inside the loop. $\Sigma^{NS}$, $\Sigma^{S}$ are defined as dimensionless parameters. Additional contributions may come from charged currents into the loop, which involves charginos and neutral sleptons.

The up quark radiative correction is analogous to the electron case but the Yukawa couplings indices and particle masses inside the loop have to be fixed in both SUSY and non-SUSY correction. In accordance to figure \ref{fig:1-loopforquarksup}, the following terms comming from the superpotential and the soft breaking potential has to be considered:
\begin{align}
    W_{Q}&\Rightarrow\hat{\mathcal{T}}_{L}\hat{\sigma}^{\prime}h_{\sigma^{\prime}}^{Tj}\hat{u}_{L}^{k c} + \hat{q}_{L}^{1}\hat{\Phi}_{2}h_{2T}^{1}\hat{\mathcal T}_{L}^{c} & V_{soft}&\Rightarrow\tilde{\lambda_{1}}\Phi_{1}^{\prime \dagger}\Phi_{2}\sigma^{\prime} + h.c. & W_{\phi}&\Rightarrow \lambda_{1}\hat{\phi}_{1}^{\prime}\hat{\phi}_{2}\hat{\sigma}^{\prime}
\end{align}

According to the figure \ref{fig:1-loopforquarksup} the SUSY and non-SUSY contributions are given in equations (\ref{NSQ}) and (\ref{SQ}), respectively:

\begin{align}
    v_{1}^{\prime}\Sigma _{1k}^{NS}(p^{2}=0)=\frac{-1}{16\pi ^2}\frac{v_{1}^{\prime}}{\sqrt{2}}\frac{\tilde{\lambda}_{1}h_{\sigma'}^{Tk}h_{2T}^{1}}{M_{T}}C_0\left(\frac{m_{h2}^{\prime}}{M_{T}},\frac{m_{\sigma}^{\prime}}{M_T}\right). \label{NSQ}
\end{align}
\begin{align}\label{SQ}
    v_{1}^{\prime}\Sigma_{1k}^{S}(p^{2}=0) = -\frac{1}{32\pi^{2}}&\frac{v_{1}^{\prime}}{\sqrt{2}}\sum_{m=1}^{8}Z_{U}^{\tilde{T}_{L}^{c}m}Z_{L}^{\tilde{T}_{L}m} \lambda_{1}h_{\sigma'}^{Tk}h_{2T}^{1}\times \\
    &\times\left[\frac{(\tilde{m}_{\sigma}^{\prime}+\tilde{m}_{h_{2}}^{\prime})^{2}}{\tilde{M}_{T_{m}}^{2}}C_{0}\left(\frac{\tilde{m}_{h1}^{\prime}}{\tilde{M}_{T_{m}}},\frac{\tilde{m}_{\sigma}^{\prime}}{\tilde{M}_{T_{m}}} \right) + \tilde{m}_{h2}^{\prime 2}B_{0}(0,\tilde{m}_{\sigma}^{\prime},\tilde{M}_{T_{m}}) + \tilde{m}_{\sigma}^{\prime 2}B_{0}(0,\tilde{m}_{h2}^{\prime},\tilde{M}_{T_{m}}) \right] \nonumber
\end{align}
\noindent
where $M_{T}$ is the $T$ exotic quark mass and $k=1,2,3.$. $\tilde{T}_{m}$ are the squark  mass eigenstates, $\tilde{m}_{T_{m}}$ the corresponding mass eigenvalue and $Z_{U}$ the associated rotation matrix which is relating the states $\tilde{T}_{L}$ and $\tilde{T}_{L}^{c}$ with the mass eigenstates. Same as the electron radiative correction, a suppression angle is considered in such a way that the rotation matrix is not written explicitly in the non-SUSY correction and likewise is assumed for SUSY contributions due to the quadratic divergences.

Finally, for the down-like quarks it is needed to write corrections which provide the down and strange quarks masses given that the mass matrix which only provides a tree level mass to the bottom quark. Hence, the Feynman rules are constructed by considering the following terms coming from the superpotential and the soft breaking potential according to figure (\ref{fig:1-loopforquarksdown}).

\begin{align}
    W_{Q}&\Rightarrow\hat{\mathcal{J}}_{L}^{a}\hat{\sigma} h_{\sigma}^{J aj}\hat{d}_{L}^{j c } + \hat{q}_{L}^{1}\hat{\Phi '}_{1}{h}_{1J}^{1a}\hat{\mathcal J}_{L}^{a\; c} + \hat{q}_{L}^{2}\hat{\phi '}_{2}{h}_{2J}^{2a}\hat{\mathcal J}_{L}^{a\; c} \\ 
    W_{\phi} &\Rightarrow \lambda_{1}\hat{\Phi}_{1}^{\prime}\hat{\Phi}_{2}\hat{\sigma}^{\prime} - \mu_{\sigma} \hat{\sigma}^{\prime}\hat{\sigma} \\
    V_{soft}& \Rightarrow \tilde{\lambda}_{2}\Phi_{2}^{\prime}\Phi_{1}\sigma 
\end{align}

Therefore, the SUSY and non-SUSY corrections can be written as:
\begin{align}
    v_{l}^{\prime}\Sigma _{lj}^{NS}(p^{2}=0)=\frac{-1}{16\pi ^2}\frac{v_{l}^{\prime}}{\sqrt{2}}\frac{\lambda(l)h_{\sigma}^{Jaj}h_{lJ}^{l}}{M_{J^{a}}}C_0\left(\frac{m_{hl}^{\prime}}{M_{J^{a}}},\frac{m_{\sigma}^{\prime}}{M_J^a}\right).
\end{align}
\begin{small}
\begin{align}
    v_{l}^{\prime}\Sigma_{lj}^{S}(p^{2}=0)& = -\frac{1}{32\pi^{2}}\frac{v_{l}^{\prime}}{\sqrt{2}}\sum_{q=1}^{10}\sum_{k=1}^{2}Z_{D}^{\tilde{J}_{L}^{ac}q}Z_{D}^{\tilde{J}_{L}^{a}q}(1-\delta_{l1}(1-Z_{L}^{\sigma k}Z_{L}^{\sigma^{\prime} k}\mu_{\sigma})) \lambda_{l}h_{\sigma}^{J^{a}j}h_{lJ}^{la}\times \\
    &\times\left[\frac{(\tilde{m}_{\sigma k}+\tilde{m}_{h_{l}}^{\prime})^{2}}{\tilde{M}_{D_{q}}^{2}}C_{0}\left(\frac{\tilde{m}_{hl}^{\prime}}{\tilde{M}_{D_{q}}},\frac{\tilde{m}_{\sigma k}}{\tilde{M}_{D_{q}}} \right) + \tilde{m}_{h_{l+1mod2}}B_{0}(0,\tilde{m}_{\sigma k},\tilde{M}_{D_{q}}) + \tilde{m}_{\sigma k}B_{0}(0,\tilde{m}_{h_{l+1mod2}},\tilde{M}_{D_{q}}) \right] \nonumber
\end{align}
\end{small}
\noindent
The module function is used in the index $l$ because when $l=1$, the scalar particle at the top of the diagram is $\phi_{2}$ but when $l=2$ the required scalar particle is $\phi_{1}$, which corresponds to a $l+1 mod(2)$ function, returning the value "1" as required.
 $\delta_{l1}$ is the usual Kronecker delta with $l=1,2$ denoting the higgs(ino) particle inside the loop, $j=1,2,3$ runs for the left-handed quarks $u_{L}^{j}$, $J_{L}^{a}$ indicates the exotic down-like quarks contribution which in this case the index $a$ can take the values $a=1,2$. $\lambda(l)$ is a function such that $\lambda(1)=\lambda_{1}$ and $\lambda(2)=\tilde{\lambda}_{2}$. Additionally, $m_{hl}$ correspond to the scalar mass particle associated with the superfields $\hat{\Phi}_{1}$, and  $\hat{\Phi}_{2}$ for $l=1,2$, respectively and $D_{q}$ are the right-handed down-like squarks mass eigenstates. When $l=1$ the radiative corrections $\Sigma_{1j}$ generates the matrix elements in the down mass matrix which produces the down quark mass, similarly the case $l=2$ produces the strange quark mass.

\section{Discussion and Conclusions}

The model studied here is the supersymmetric extension of the three families $U(1)_{X}$ model \cite{nosusybib}. The SUSY extension requires four Higgs doublets and four scalar singlets in order to not induce chiral anomalies and giving mass to quarks, charged leptons and neutrinos. Additionally, the singlets generate the mass for exotic fermions and break the $U(1)_{X}$ gauge symmetry. An interesting prediction of this extension is that there are tree level flavor changing neutral currents.

In both versions of the model, SUSY and non SUSY, the lightest particles (electron and up, down and strange quarks) are massless at tree level. However, in the supersymmetric model they acquire a mass value via radiative corrections through inert singlets into the loop after $U(1)_{X}$ and electroweak symmetry breaking.

By implementing a seesaw mechanism among the singlet and doublet Higgs fields, together with the D-terms corresponding to the $U(1)_{X}$, a $\Delta m^2_h$ is found at tree level and it turns out to be at the order of the MSSM contribution. The lightest scalar particle is identified as the Higgs boson and its mass is obtained at the order of $125$ GeV. In fact, we  show in figures (\ref{fig:plotvpvsgx}) and (\ref{fig:plotv2vsgx}) the region in the parameter space $v_{1}'$ vs $g_{X}$ and $v_{2}$ vs $g_{X}$ which is compatible with a $125.3\text{ GeV}$ Higgs boson at 95\%  of C.L., where the VEVs $v_{1}$ and $v_{2}'$ are fixed around the mass of the top quark and the bottom quark, respectively. As a result, we found that the coupling constant $g_{X}$, regarding to the $U(1)_{X}$ symmetry, takes values between $0.63$ and $1$ for $112<v_2(\text{GeV})<180$ and $0<v'_1(\text{GeV})<81$. Thus, values below $0.63$ for $g_{X}$ are excluded. Even more, if we take the LHC exclusion bounds on dilepton production, which gave $m_{Z'}>8$TeV, it implies from the expression for this gauge boson mass (eq. (\ref{masadelnuevoz})) that the VEVs $v_\chi \approx v'_\chi$ should be greater than $37$ TeV.

 The parameter space $\cos 2\Tilde{\beta}$ vs $g_{X}$ was explored and negative values are found for $\cos 2\Tilde{\beta}$ because non-primed VEV happened to be greater than the primed ones. This behavior lies in the fact that top quark mass ($\approx v_1$) is bigger than the bottom quark mass ($\approx v'_2$).  On the other hand, in the mass expression for $m_h$, eq. \ref{Higgsobservedmass}, there are quadratic differences between non-primed and primed VEVs so $v_{1},v_{2}> v_{1}',v_{2}'$ is preferred. Thus the allowed region is $0.38<\cos \Tilde{\beta}<1$.

Last but not less important, the model also predicts five CP-even, four CP-odd and three charged Higgs particles with a mass above the TeV scale. The would-be Goldstone bosons corresponding to $Z_\mu$, $W_\mu^\pm$ and $Z'_\mu$ are also found. 

\appendix

\section{Scheme for obtaining the scalar particle mass expressions}

For the CP-even particles an additional first step was made, which is to perform a seesaw like rotation, taking into account that the mixing in $M_{h\xi}$ are small compared to the ones in $M_{\xi \xi}$. With that approximation, the matrix $M_{h}$ is transformed to a block diagonal form:
\begin{equation}
M_{h}\rightarrow (\mathbb{V}^{h}_{\mathrm{SS}})^{T}M_{h}\mathbb{V}^{h}_{\mathrm{SS}} \approx \begin{pmatrix}
    \Tilde{M}_{hh} & 0  \\
    0 & M_{\xi \xi}  
    \end{pmatrix}, 
\end{equation}
 where $\Tilde{M}_{hh}=M_{hh}-M_{h \xi} M_{\xi \xi}^{-1} M_{h \xi}^{T}$. $\mathbb{V}^{h}_{\mathrm{SS}}$ rotates the matrix,
\begin{equation}
\mathbb{V}^{h}_{\mathrm{SS}}=\begin{pmatrix}
1&\Theta^{h\dagger}\\
-\Theta^{h}&1,
\end{pmatrix}    
\end{equation}
with $\Theta^{h}=M_{h\xi}(M_{\xi\xi})^{-1}$. The heavy remaining block component, $M_{\xi \xi}$, is a block diagonal $4\times 4$ matrix, and therefore it's eigenvalues can be obtained trivially.


\bibliographystyle{ieeetr}

\begin{thebibliography}{}

\end{thebibliography}


\vspace{0.1mm}

\begin{thebibliography}{99}

\bibitem{SMbib} S.L. Glashow, Nucl. Phys. 22, 579 (1961); A. Salam, Elementary Particle Physics (Nobel Symp. No. 8), edited by N. Svartholm (Almquist and Wiksells, Stockholm, 1968), p. 367; S. Weinberg, Phys. Rev. Lett. 19, 1264 (1967).

\bibitem{Higgsobservationbib} G. Aad, et al., Phys. Lett. B716, 1 (2012); S. Chatrchyan, et al., Phys. Lett. B716, 30 (2012).

\bibitem{muproblembib}  J. E. Kim and H. Nilles, Phys. Lett. B138, 150 (1984).

\bibitem{NMSSMbib} T. Elliott, S. F. King, and P. L. White, Phys. Rev. D49, 2435 (1994); D. Miller, R. Nevzorov, and P. Zerwas, Nucl. Phys. B681, 3 (2004); S. F. King and P. L. White, Phys. Rev. D53, 4049 (1996); B. A. Dobrescu and K. T. Matchev, JHEP 09, 031 (2000); S. King, M. Mühlleitner, and R. Nevzorov, Nucl. Phys. B860, 207 (2012).

\bibitem{HiggsmassMSSMbib} H. E. Haber and R. Hempfling, Phys. Rev. Lett. 66,  1815 (1991); M. Yamaguchi, T. Yanagida, and Y. Okada, Prog. of Theor. Phys. 85, 01 (1991); J. Ellis, G. Ridolfi, and F. Zwirner, Phys. Lett. B262, 477 (1991); M. Carena, J. Espinosa, M. Quirós, and C. Wagner, Phys. Lett. B355, 209 (1995); M. Carena, M. Quirós, and C. Wagner, Nucl. Phys. B461, 407 (1996); H. E. Haber, R. Hempfling, and A. H. Hoang Z. Phys. C 75, 539 (1997);  S. Heinemeyer, W. Hollik, and G. Weiglein, Phys. Rev. D58, 091701 (1998); S. Heinemeyer, W. Hollik, and G. Weiglein, Phys. Lett. B440, 296 (1998); S. Heinemeyer, W. Hollik, and G. Weiglein, EPJ C 9, 343 (1999); J. R. Espinosa and R.-J. Zhang, JHEP 2000, 026 (2000); M. Carena, et al., Nucl. Phys. B580, 29 (2000); S. P. Martin, Phys. Rev. D67, 095012 (2003).

\bibitem{SeesawHiggsMass} P. Batra, A. Delgado, D. E. Kaplan, and T. M. Tait, JHEP 2004, 043 (2004); P. Batra, A. Delgado, D. E. Kaplan, and T. M. Tait, JHEP 2004, 032 (2004); Bellazzini, C. Csáki, A. Delgado, and A. Weiler, Phys. Rev. D79, 095003 (2009).

\bibitem{USSMbib} H. Goldberg, Phys. Rev. Lett. 50, 1419 (1983); J. Ellis, D. Nanopoulos, and K. Tamvakis, Phys. Lett. B121, 123 (1983); R. Harnik, D. T. Larson, H. Murayama, and M. Thormeier, Nucl. Phys. B706, 372 (2005); H.-C. Cheng, B. A. Dobrescu, and K. T. Matchev,Phys. Lett. B439, 301 (1998); M. Aoki and N. Oshimo, Phys. Rev. Lett. 84, 5269 (2000).

\bibitem{DMSUSYbib} H. Pagels and J. R. Primack, Phys. Rev. Lett. 48, 223 (1982); H. Goldberg, Phys. Rev. Lett. 50, 1419 (1983).

\bibitem{FCNCbib} S. King, S. Moretti, and R. Nevzorov, Phys. Letters B634, 278 (2006); S. F. King, S. Moretti, and R. Nevzorov,Phys. Rev. D73, 035009 (2006).

\bibitem{dataparticle} C. Patrignani et al. [Particle Data Group], Chin. Phys. C40, 100001 (2016). doi:10.1088/1674-1137/40/10/100001.

\bibitem{nosusybib} S. Mantilla, R. Martinez, and F. Ochoa, Phys. Rev. D95, 095037 (2017).

\bibitem{KM} Makoto Kobayashi, Toshihide Maskawa, Prog. Theor. Phys. 49, 652  (1973).

\bibitem{CKMPMNS} M.C. Gonzalez-Garcia, M. Maltoni, and T. Schwetz., Nucl. Phys. B908, 199 (2016);  H. Fritzsch and S. Zhou, Phys. Lett. B 718, 1457 (2013).

\bibitem{jerarquia}H. Georgi, Nucl. Phys. B156, 126 (1979); J.E. Kim, Phys. Rev. Lett. 45, 1916 (1980); P.H. Frampton, Phys. Lett. B89, 352 (1980); S. Barr, Phys. Rev. D21, 1424 (1980); R. Barbieri and D.V. Nanopoulos, Phys. Lett. B91, 369 (1980); L.E. Ib ifiez, Nucl. Phys. B193, 317 (1982); F. Wilczek and A. Zee, Phys. Rev. Lett. 42, 421 (1979); A. Davidson, K.C. Wali and P.D. Manheim, Phys. Rev. Lett. 45, 1135 (1980); H. Fritzsch, Phys. Lett. B70, 436 (1977); H. Fritzsch, Phys. Lett. B73, 317 (1978); H. Fritzsch, Nucl. Phys. B155, 189 (1979); H. Fritzsch and J. Plankl, Phys.Rev. D35, 1732 (1987); C. D. Froggatt and H. B. Nielsen, Nucl. Phys. B147, 277 (1979);  K.S. Babu and Ernest Ma, Phys. Rev. Lett. 61, 674 (1988); Ernest Ma, Phys. Rev. Lett. 62, 1228 (1989); Ernest Ma, Phys. Rev. Lett. 63, 104 (1989); Ernest Ma, Phys. Rev. Lett. 64, 2866 (1990); Ernest Ma, Phys.Rev. D80, 013013 (2009); S. M. Barr, Phys. Rev. D21, 1424, (1980); B. Balakrishna, Phys. Rev. Lett. 60, 1602 (1988); B. Balakrishna, A. Kagan, and R. Mohapatra, Phys. Lett. B205, 345 (1988);  Y. Koide, Phys. Rev. D28, 252 (1983);  Z.Z. Xing, Nuovo Cimento A (1971-1996) 109, 115 (1996); N. Arkani-Hamed and M. Schmaltz, Phys. Rev. D61, 033005 (2000); K. Yoshioka, Mod. Phys. Lett. A15, 29 (2000); E. A. Mirabelli and M. Schmaltz, Phys. Rev. D61, 113011 (2000); L. Randall and R. Sundrum, Phys. Rev. Lett. 83, 3370 (1999).

\bibitem{anzat}H. Fritzsch, Phys. Lett. B70, 436 (1977) ; B73, 317 (1978); Nucl. Phys. B155, 189 (1979); T.P. Cheng and M. Sher, Phys. Rev. D35, 3484 (1987); D. DU and Z. Z. Xing, Phys. Rev. D48, 2349 (1993); H. Fritzsch and Z. Z. Xing, Phys. Lett. 555, 63 (2003); J.L. Diaz-Cruz, R. Noriega Papaqui and A. Rosado, Phys. Rev. D71, 015014 (2005); K. Matsuda and H. Nishiura, Phys. Rev. D74, 033014 (2006); A. Carcamo, R. Martinez and J.-A. Rodriguez, Eur. Phys. J. C50, 935 (2007); Fritzsch, Harald. Phys. Lett. B73, 317 (1978); Fukugita, Masataka, Morimitsu Tanimoto, and Tsutomu Yanagida. Prog. of Theor. Phys. 89, 263 (1993); Fukugita, M., and Tsutomu Yanagida. Phys. Lett. B174, 45 (1986).

\bibitem{corrienteadicional}P. Langacker and M. Plumacher, Phys. Rev. D62, 013006 (2000); K. Leroux and D. London, Phys. Lett. B526, 97 (2002); S. Baek, J. H. Jeon, and C. S. Kim, Phys. Lett. B641, 183 (2006; E. Ma, Phys. Lett. B380, 286 (1996); V. Barger, P. Langacker and H.-S. Lee, Phys. Rev. D67, 075009 (2003), S.F. King., S. Moretti and R. Nevzorov, Phys. Rev. D73, 035009 (2006). 23; T. Hur, H.-S. Lee, and S. Nasri, Phys. Rev. D77, 015008 (2008); G. Belanger, A. Pukhov, and G. Servant, JCAP 0801, 009 (2008); Lorenzo Basso, Stefano Moretti, Giovanni Marco Pruna JHEP 1108, 122 (2011); Elena Accomando, Claudio Coriano, Luigi Delle Rose, Juri Fiaschi, Carlo Marzo and Stefano Moretti, JHEP 1607, 086 (2016); Elena Accomando, Luigi Delle Rose, Stefano Moretti, Emmanuel Olaiya, Claire H. Shepherd-Themistocleous, JHEP 1704, 081 (2017);  P. Langacker, Rev, Mod. Phys. 81, 1199 (2009) [hep-ph/0801.1345]; A. Leike, Phys. Rep. 317, 143 (1999); J. Erler, P. Langacker, and T. J. Li, Phys. Rev. D 66, 015002 (2002); S. Hesselbach, F. Franke, and H. Fraas, Eur. Phys. J. C 23, 149 (2002);  Martinez, R., Nisperuza, J., Ochoa, F., and Rubio, J. Physical Review, D 90  (2014), 095004; Martinez, R., Nisperuza, J., Ochoa, F., Rubio, J. P., and Sierra, C. F. Phys. Rev. D92, 035016 (2015) ;  Martinez, R., Ochoa, F., JHEP 05, 113 (2016); R. Martınez, J. Nisperuza, F. Ochoa, and J. P. Rubio. Phys. Rev. D 89, 056008; C.E Díaz, S.F Mantilla and R. Martínez, Phys. Rev. D98, 015038 (2018); A.~Das, S.~Oda, N.~Okada and D.~s.~Takahashi, Phys.\ Rev.\ D{\bf 93}, 115038 (2016); A.~Das, N.~Okada and D.~Raut, Phys.\ Rev.\ D{\bf 97}, 115023 (2018); A.~Das, N.~Okada and D.~Raut, Eur.\ Phys.\ J.\ C{\bf 78}, 696 (2018).

\bibitem{dibosonproduction} P. Osland, A.A. Pankov, and A.V. Tsytrinov, Phys. Rev. D 96, 055040 (2017).

\bibitem{CMSdataondiboson} The CMS Collaboration [CMS Collaboration], CMS-PAS-B2G-17-001.

\bibitem{ATLASdataondiboson} The ATLAS collaboration [ATLAS Collaboration], ATLAS-CONF-2017-051.

\bibitem{ATLASdataondilepton} The ATLAS collaboration, Aaboud, M., Aad, G. et al. J. High Energ. Phys., 182 (2017). 
\bibitem{dataonexotic} Serkin, Leonid, arXiv:1901.01765 (2019) 
\bibitem{exposcilacion} Davis Jr, R., Harmer, D. S., Hoffman, K. C., Phys. Rev. Lett. 20, 1205 (1968); Abdurashitov, J. N., et. al.. Phys. Rev. C80, 015807 (2009); Kaether, F., Hampel, W., Heusser, G., Kiko, J., Kirsten, T. Phys. Lett. B685, 47 (2010); Cleveland, B. T., et. al.. The Astrophysical Journal 496, 505 (1968); Aharmim, B., et. al.. Phys. Rev. C88, 025501 (2013); Borexino Collaboration. Bellini, G., et. al., Phys. Rev. D82, 033006 (2010); Borexino Collaboration. Bellini, G., et. al., Nature 512, 383 (2014);  Hosaka, J., et. al., Phys. Rev. D73, 112001 (2006); Hosaka, J., et. al., Phys. Rev. D74, 032002 (2006); Cravens, J. P., Abe, K., et. al., Phys. Rev. D78, 032002 (2008);  Aartsen, M. G., et al. Phys. Rev. D91, 072004 (2015);  Gando, A., et al. Phys. Rev. D83, 052002 (2011);  Apollonio, Marco, et al. Phys. Lett. B466, 415 (1999); Piepke, A. et al. [Palo Verde Collaboration]. Progress in Particle and Nuclear Physics 48(1), 113 (2002); F. P. An et al. [Daya Bay Collaboration], Phys. Rev. Lett. 115, 111802 (2015);  S. B. Kim et al. [RENO Collaboration], Nucl. Phys. B908, 94 (2016); Kopp, J. et al. JHEP, 2013(5), 1 (2013); Adamson, P. et al. Phys. Rev. Lett. 110(25), 251801 (2013); Adamson, P. et al. Phys. Rev Lett. 110(17), 171801 (2013);  K. Abe et al. [T2K Collaboration], Phys. Rev. Lett. 112, 061802 (2014) ; K. Iwamoto [T2K Collaboration], ICHEP2016, Chicago, 06 August, 2016; J. Bian [NO$\nu$A Collaboration], ICHEP2016, Chicago, 06 August, 2016; P. Adamson et al. [NO$\nu$A Collaboration], Phys. Rev. Lett. 116, 151806 (2016).


\bibitem{ISSbib} R. N. Mohapatra, Phys. Rev. Lett. 56, 561 (1986); R. N. Mohapatra and J. W. Valle, Phys. Rev. D34, 1642 (1986); E. Catano, R. Martinez, and F. Ochoa, Phys. Rev. D86, 073015 (2012); P. S. Bhupal Dev and A. Pilaftsis, Phys. Rev. D86, 113001 (2012).

\bibitem{Ferraribib} M. Abramowitz and I. Stegun, EPJ C, 17 (1972).

\bibitem{RadCorrSusybib} A. Crivellin and U. Nierste, Phys. Rev. D81, 095007 (2010); J. Foster, K. ichi Okumura, and L. Roszkowski, JHEP 2005, 094 (2005); A. Crivellin and U. Nierste, Phys. Rev. D79,  035018 (2009); M. Carena, D. Garcia, U. Nierste, and C. E. Wagner, Nucl. Phys. B577, 88 (2000).

\bibitem{Feynmannrulesbib} J. Rosiek, Phys. Rev. D41, 3464 (1990).

\end{thebibliography}

\end{document}